\documentclass[sigconf]{acmart}
\makeatletter
\def\@ACM@checkaffil{
    \if@ACM@instpresent\else
    \ClassWarningNoLine{\@classname}{No institution present for an affiliation}%
    \fi
    \if@ACM@citypresent\else
    \ClassWarningNoLine{\@classname}{No city present for an affiliation}%
    \fi
    \if@ACM@countrypresent\else
        \ClassWarningNoLine{\@classname}{No country present for an affiliation}%
    \fi
}
\makeatother

\usepackage{hyperref}

\usepackage{amsmath,amsfonts}
\usepackage{algorithmic}
\usepackage{tikz}
\usepackage{graphicx}
\usepackage{textcomp}
\usepackage{xcolor}
\usepackage{booktabs}
\usepackage{pifont}
\usepackage{siunitx}
\usepackage{tabularray}
\usepackage{adjustbox}
\usepackage{amsmath}
\usepackage{algorithm2e}
\usepackage{multirow}
\usepackage{fancyhdr}
\usepackage{float}
\usepackage{setspace}
\usepackage{listings}
\usepackage{balance}
\usepackage{wrapfig}
\usepackage{marginnote}
\usepackage{caption}
\usepackage{boldline} 
\usepackage{colortbl} 
\usepackage{floatflt}
\usepackage{listing}
\usepackage{eso-pic}
\usepackage{titlesec}
\usepackage{noindentafter}
\usepackage{datetime}
\usepackage[htt]{hyphenat}

\AtBeginDocument{%
  \providecommand\BibTeX{{%
    \normalfont B\kern-0.5em{\scshape i\kern-0.25em b}\kern-0.8em\TeX}}}

\usepackage{xspace}
\usepackage{color}
\usepackage{listings}
\usepackage{xcolor}

\definecolor{darkspringgreen}{rgb}{0.09, 0.45, 0.27}
\definecolor{denim}{rgb}{0.08, 0.38, 0.74}
\definecolor{darkolivegreen}{rgb}{0.33, 0.42, 0.18}
\definecolor{tangerine}{rgb}{0.95, 0.52, 0.0}
\definecolor{mahogany}{rgb}{0.75, 0.25, 0.0}

\newcommand{\shellcmd}[1]{\indent\indent\texttt{\footnotesize\$ #1}}

\definecolor{uglyyellow}{rgb}{0.99, 0.93, 0.0}
\definecolor{nbs}{rgb}{0.35, 0.31, 0.81}
\newcommand{\nb}[1]{\textcolor{black}{#1}}

\definecolor{darkred}{rgb}{0.75, 0.1, 0.1}
\newcommand{\dbb}[1]{\textcolor{black}{#1}}

\usepackage{tikz}

\newcommand{\squishlist}{
 \begin{list}{$\circ$}
  { \setlength{\itemsep}{0pt}
     \setlength{\parsep}{0pt}
     \setlength{\topsep}{3pt}
     \setlength{\partopsep}{0pt}
     \setlength{\leftmargin}{1em}
     \setlength{\labelwidth}{1em}
     \setlength{\labelsep}{0.5em} } }

\newcommand{\squishend}{
  \end{list}  }

\newcommand*\circled[1]{\tikz[baseline=(char.base)]{\node[shape=circle,fill,inner sep=0.5pt] (char) {\textcolor{white}{#1}};}}
\newcommand*\circledwhite[1]{\tikz[baseline=(char.base)]{\node[shape=circle, inner sep=0.5pt, draw=black, fill=white, text=black] (char) {#1};}}

\usepackage[colorinlistoftodos,prependcaption,textsize=scriptsize]{todonotes}

\usepackage[bottom]{footmisc}
\usepackage{dblfloatfix}
\usepackage{array}
 \usepackage{booktabs}

\newcommand\head[1]{{\noindent\textbf{#1}.}}
\definecolor{ufogreen}{rgb}{0.1, 0.6, 0.4}
\definecolor{ufogreen}{rgb}{0.1, 0.6, 0.4}

\newcommand{\system}{Victima\xspace}
\newcommand{\speedupoverbaseline}{7.4} 
\newcommand{\speedupoverpomtlb}{6.2} 
\newcommand{\speedupoverbaselinevirt}{28.7} 
\newcommand{\speedupoverispvirt}{4.9} 
\newcommand{\speedupoverpomtlbvirt}{20.1} 
\newcommand{\latencyoverbaselinenative}{22}
\newcommand{\latencyoverbaselinevirt}{60}

\newcommand{\area}{0.04} 
\newcommand{\power}{0.08} 

\usepackage{listings}
\usepackage{chngcntr}
\usepackage{xcolor}
\usepackage[T1]{fontenc} 
\usepackage{zi4} 
\usepackage{caption} 
\usepackage{float} 

\lstdefinestyle{custompseudocode}
{
    language=Java,
    basicstyle=\footnotesize\ttfamily,
    commentstyle=\ttfamily\itshape\color{gray},
    stringstyle=\ttfamily,
    showstringspaces=false,
    breaklines=true,
    belowcaptionskip=0.3\baselineskip, 
    frameround=ffff,
    frame=single,
    rulecolor=\color{black},
    tabsize=1,
    keywordstyle=\color{red}\bfseries,
    columns=fullflexible,
    morekeywords={public, class}
    morekeywords={function}, 
    keywordstyle=[2]\bfseries, 
    escapeinside={*@}{@*}, 
}
\lstdefinestyle{customc}{
  belowcaptionskip=1\baselineskip,
  breaklines=true,
  frame=single,
  xleftmargin=\parindent,
  language=C,
  showstringspaces=false,
  basicstyle=\footnotesize\ttfamily,
  keywordstyle=\bfseries\color{green!40!black},
  commentstyle=\itshape\color{tangerine},
  identifierstyle=\color{black},
  stringstyle=\color{orange},
  numbers=left,
  stepnumber=1,
  numbersep=-5pt,
  escapeinside={*@}{@*}, 
}

\DeclareCaptionFormat{listing}{
  \rule{\linewidth}{0.5pt}
  \vspace{-1mm}
  \textbf{#1#2: }#3
  \vspace{-1mm}
  \rule{\linewidth}{0.5pt}
  \vspace{-7mm}
}

\definecolor{cadmiumgreen}{rgb}{0.0, 0.42, 0.24}
\definecolor{chocolate}{rgb}{0.92, 0.41, 0.12}
\definecolor{burgundy}{rgb}{0.5, 0.0, 0.13}
\definecolor{darkmagenta}{rgb}{0.55, 0.0, 0.55}
\definecolor{darkblue}{rgb}{0.0, 0.5, 1.0}
\newcommand\konkanelloreva[1]{\textcolor{black}{#1}}
\newcommand\konkanellorevb[1]{\textcolor{black}{#1}}

\newcommand\konkanellorevd[1]{\textcolor{black}{#1}}
\newcommand\konkanelloreve[1]{\textcolor{black}{#1}}

\newcommand\konrev[1]{\textcolor{black}{#1}}
\newcommand\konreva[1]{\textcolor{black}{#1}}
\newcommand\konrevb[1]{\textcolor{black}{#1}}
\newcommand\konrevc[1]{\textcolor{black}{#1}}
\newcommand\konrevd[1]{\textcolor{black}{#1}}
\newcommand\konreve[1]{\textcolor{black}{#1}}
\newcommand\konrevs[1]{\textcolor{black}{#1}}
\newcommand\konrevf[1]{\textcolor{black}{#1}}

\copyrightyear{2023}
\acmYear{2023}
\setcopyright{acmlicensed}\acmConference[MICRO '23]{56th Annual IEEE/ACM International Symposium on Microarchitecture}{October 28-November 1, 2023}{Toronto, ON, Canada}
\acmBooktitle{56th Annual IEEE/ACM International Symposium on Microarchitecture (MICRO '23), October 28-November 1, 2023, Toronto, ON, Canada}
\acmPrice{15.00}
\acmDOI{10.1145/3613424.3614276}
\acmISBN{979-8-4007-0329-4/23/10}
 \newif\ifcameraready
\camerareadytrue

\newif\ifarxiv
\arxivtrue

\ifarxiv
\usepackage{calc}
\fancypagestyle{firstpage}
{
    \fancyhead{}
    \begin{tikzpicture}[remember picture,overlay]
    \node [xshift=134mm,yshift=-10mm]
    at (current page.north west) {\href{https://www.acm.org/publications/policies/artifact-review-and-badging-current}{\includegraphics[width=1.8cm]{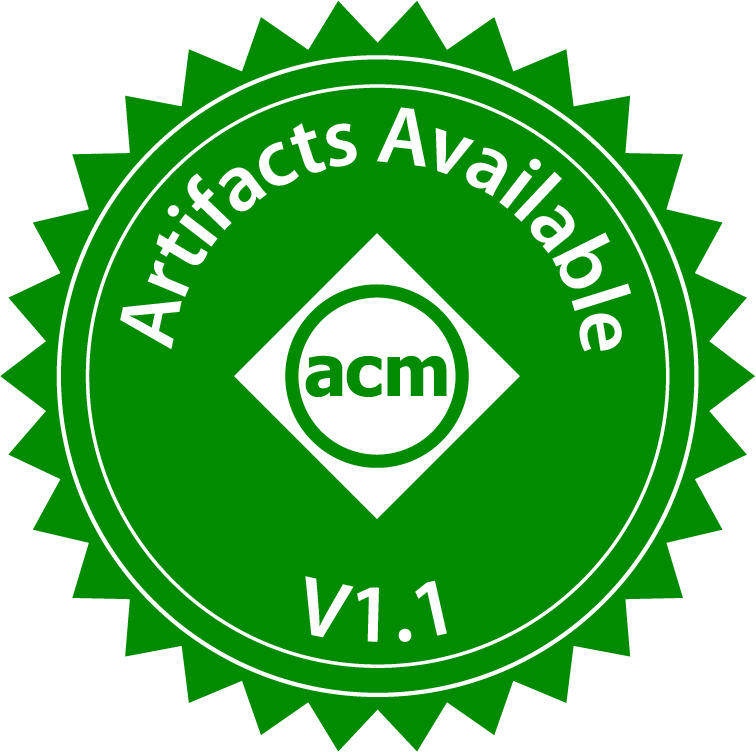}}} ;
    \node [xshift=153mm,yshift=-10mm]
    at (current page.north west) {\href{https://www.acm.org/publications/policies/artifact-review-and-badging-current}{\includegraphics[width=1.8cm]{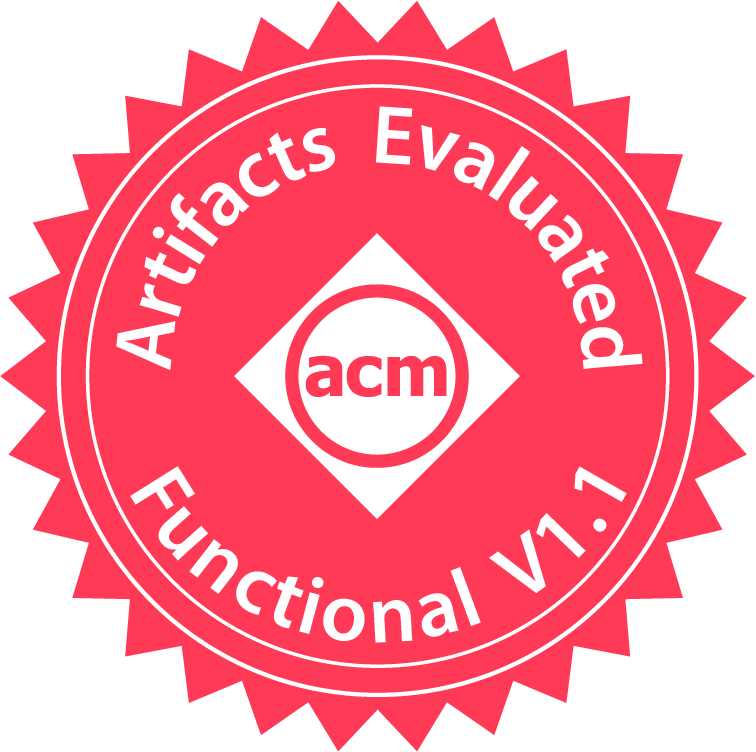}}} ;
    \node [xshift=172mm,yshift=-10mm]
    at (current page.north west) {\href{https://www.acm.org/publications/policies/artifact-review-and-badging-current}{\includegraphics[width=1.8cm]{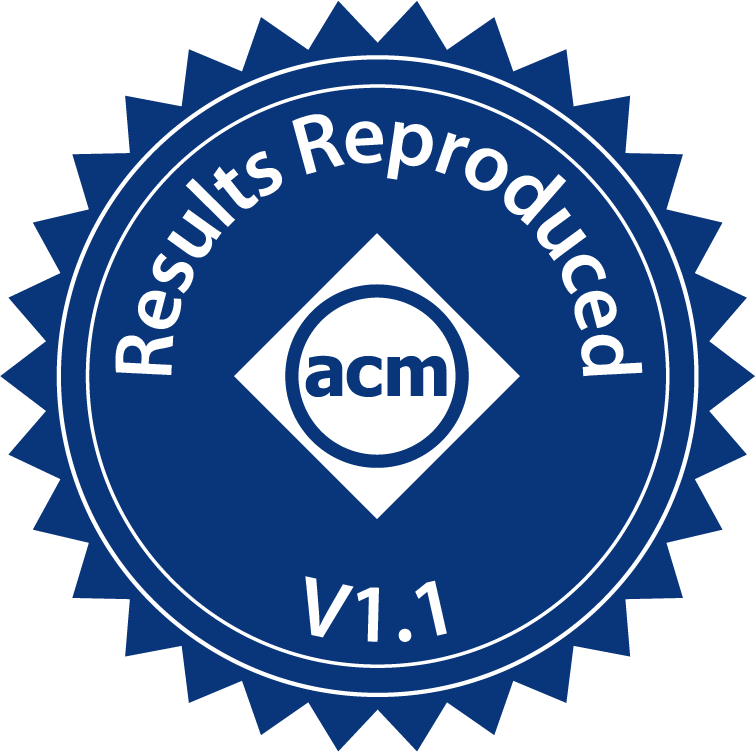}}} ;
    \node [xshift=191mm,yshift=-10mm]
    at (current page.north west) {\href{https://www.acm.org/publications/policies/artifact-review-and-badging-current}{\includegraphics[width=1.8cm]{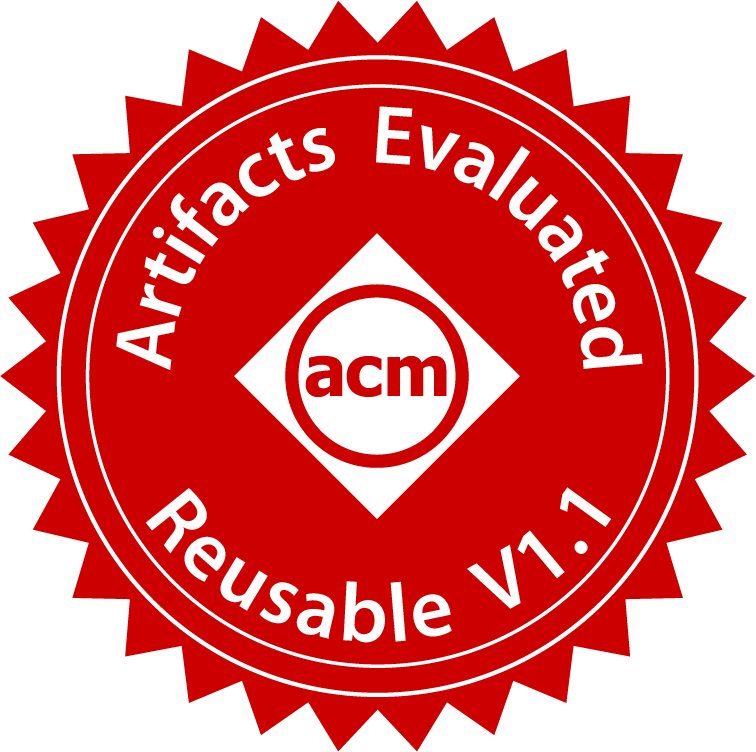}}} ;
    \end{tikzpicture}
  
  \pagenumbering{arabic}
  \fancyfoot[C]{\large\thepage}
}
\fi

\newcommand\VMcharacterization{~\cite{vm2,karakostas_characterIISWC,
isca2010-barr-trancache,5-levelpaging,contiguitas2023,radiantISMM21,bhattacharjeePACT2009,devirtualizingASPLOS2018,hash_dont_cache,virtualizationimplication,vm29}}

\newcommand\VMlargepages{~\cite{park2020perforated,guvenilir2020tailored,ingensOSDI2016,talluriISCA1992,panwar2018making,panwar2019hawkeye,tridentMICRO2021,pham2015,mosaic2017MICRO,promotionHPCA2001,shadowpageISCA1998,duHPCA2015,vm42,vm43,partialMICRO2020,ganapathy98}}

\newcommand\VMcontiguity{~\cite{vm2,rlbenergyHPCA2016,translationranger2019,karakostas2015,chloe2020,hybridtlbISCA2017,flexpointerTACO2023,contiguitas2023,vm6}}

\newcommand\VMtlblthree{~\cite{sharedl3tlbISCA2011,distlltlbMICRO2018,gpustealing}}

\newcommand\VMcontextswitch{~\cite{kaffesHotOS2021,XPC}}

\newcommand\VMpwcs{~\cite{isca2010-barr-trancache,vm10,esteve14}}

\newcommand\VMpagetable{~\cite{haria2018, flataAsplos2022,elastic-cuckoo-asplos20,mehtJovanHPCA2023,hash_dont_cache,mosaicpagesASPLOS2023,kanellopoulosMICRO2023utopia,nearmemoryPact17,impicaICCD2016,mitosis-asplos20,compendiaISMM2021,Alam2017DoItYourselfVM}}

\newcommand\VMvirtualized{~\cite{vm25,vm35,pham2015,pham2015tr,vm11,virtcoherenceISCA2017,babelfish,margaritov2021ptemagnet,virtcoherenceISCA2017,panwar2021fast}}

\newcommand\VMvirtualcaching{~\cite{kaxiras2013,seesawISCA2018,basu2012, cekleov1997a, wood1986,coherencyvirtualASPLOS1987,consistencyvirtualASPLOS1992,virtualcacheISCA1989}}

\newcommand\VMintermediate{~\cite{enigma,midgard,vbi,powerpc2003}}

\newcommand\VMtlbprefetching{~\cite{vavouliotis2021,morriganMICRO2021,margaritov2019prefetched,kandiraju2002going,saulsbury2000recency,Bala1994SoftwarePA}}

\newcommand\VMtlbreplacementpolicy{~\cite{deadTLBHPCA2021,chirpMICRO2020}}

\newcommand\VMtlball{~\cite{chirpMICRO2020,papadopoulou2015,latr,juan97,onlinesuperpagepromotionISCA1995,compilerdtlbISPASS2004,vm38,vm39,wood1986,skewedTLB,vavouliotis2021,morriganMICRO2021,margaritov2019prefetched,kandiraju2002going,saulsbury2000recency,Bala1994SoftwarePA,isca2010-barr-trancache,vm10,sharedl3tlbISCA2011,distlltlbMICRO2018}}

\newcommand\VMsoftwareTLB{~\cite{pomtlbISCA2017,csaltMICRO2017,softwareTLBNAS2013,softwareTLBISCA2013,uhlig94,bruceMMU1998,softcontrolcachesISCA1986,Nagle1993DesignTF,Bala1994SoftwarePA}}

\newcommand\VMold{~\cite{ieemicro2018-Bhattacharjee-tempo,hand1999,old_vm1,old_vm2,old_vm3,old_vm4,old_vm5,old_vm6,old_vm7,denning1970,ahearn1973,goldberg1974survey,bruceMMU1998,smith,wood1986,chen1992simulation,koldinger1992,lindstrom1995,jacob1998,avm,translationmanagementISCA1993,interactionASPLOS1991,multics}}

\begin{document}
\counterwithout{lstlisting}{subsection}

 \title{Victima: Drastically Increasing Address Translation Reach \\ by Leveraging Underutilized Cache Resources }

\author{
  Konstantinos Kanellopoulos\textsuperscript{1}\quad
  Hong Chul Nam\textsuperscript{1}\quad
  F. Nisa Bostanci\textsuperscript{1}\quad
  Rahul Bera\textsuperscript{1} \\
  Mohammad Sadrosadati\textsuperscript{1}\quad
  Rakesh Kumar\textsuperscript{2}\quad
  Davide Basilio Bartolini\textsuperscript{3}\quad
  Onur Mutlu\textsuperscript{1}
  \\
}
\affiliation{%
  \vspace{0.7em}
  \institution{\textsuperscript{1}ETH Zürich\quad \textsuperscript{2}Norwegian University of Science and Technology\quad  \textsuperscript{3}Huawei Zurich Research Center}
}
\email{}

 \renewcommand{\shortauthors}{Kanellopoulos et al.}
 \renewcommand{\shorttitle}{Victima: Drastically Increasing Address Translation Reach\\ by Leveraging Underutilized Cache Resources}

\renewcommand{\authors}{ Konstantinos Kanellopoulos,
Hong Chul Nam,
F. Nisa Bostanci,
Rahul Bera,
Mohammad Sadrosadati,
Rakesh Kumar,
Davide Basilio Bartolini,
Onur Mutlu}

\begin{abstract}
    Address translation is a performance bottleneck in data-intensive workloads due to \konrev{large} datasets and irregular access patterns \konreva{that} lead to frequent  high-latency page table walks (PTW\konrev{s}).
   \konrev{PTWs can be \konreva{reduced by} using (i) large hardware TLBs or (ii) large software-managed TLBs. 
    Unfortunately, both solutions have significant drawbacks: increased access latency, power and area \konreva{(for hardware TLBs)}, and costly memory accesses, the need for large contiguous memory blocks, and complex OS modifications  \konreva{(for software-managed TLBs)}.}

    We present \system, a new \textit{software-transparent} mechanism that drastically increases the translation reach of the processor by leveraging the \konrev{underutilized} resources of the cache hierarchy. 
    The \textbf{key idea} of \system is to repurpose L2 cache blocks to store clusters of TLB entries, \konrev{thereby} providing an additional low-latency and high-capacity component that \konrev{backs up the last-level TLB} and \konreva{thus} reduces PTWs.
    \system \konreva{has} two main components\konrev{.} First, a PTW cost predictor (PTW-CP) \konreva{identifies} costly-to-translate addresses based on the frequency and cost of the PTWs \konreva{they lead to}. 
    \konreva{\konrevb{Leveraging} the PTW-CP, Victima uses the valuable cache space only for TLB entries that correspond to costly-to-translate pages, reducing the impact on cached application data.} Second, 
    a TLB-aware cache replacement policy \konrev{prioritizes keeping TLB entries in the cache hierarchy} \konreva{by considering (i) the translation pressure (\konreva{e.g.}, \konrevb{last-level TLB miss rate}) and (ii) the reuse \konrevb{characteristics} of the TLB entries.}

    Our evaluation results show that in native (virtualized) execution environments \system improves \konreva{average} end-to-end application performance  by \speedupoverbaseline\% (\speedupoverbaselinevirt\%) 
    over the baseline four-level radix-tree-based page table design and by \speedupoverpomtlb\% (\speedupoverpomtlbvirt\%) \konrev{over a} state-of-the-art software-managed TLB, \konreva{across 11 diverse data-intensive workloads}.
    Victima delivers \konreva{similar} performance as a system that employs an optimistic 128K-entry L2 TLB, while avoiding the associated area and power overheads.
    \system  (i) is effective in both native and virtualized environments,  
    (ii) is completely transparent to \konreva{application and system} software, (iii) \konrev{unlike large software-managed TLBs}, does not require contiguous physical allocations, (iv) is \konreva{compatible} with modern large page mechanisms 
    and (iv) incurs \konreva{very small} area and power overhead\konrev{s} of $\area\%$ and $\power\%$, respectively, \konrev{on} a modern high-end CPU. The source code of Victima is freely available
    at \textcolor{blue}{\url{https://github.com/CMU-SAFARI/Victima}}.

\end{abstract}

%

\maketitle
\thispagestyle{firstpage}

\section{Introduction}
\label{sec:intro}

Address translation is a significant performance bottleneck in modern data-intensive workloads\VMcharacterization. 
To enable fast address translation, modern processors employ a two-level translation look-aside buffer (TLB) hierarchy that caches recently used virtual-to-physical address translations.
However, with the \konreva{very large} data footprint\konreva{s} of modern workloads, the last-level TLB (L2 TLB) experiences high miss \konreva{rate (misses per kilo instructions; MPKI)}, leading to high-latency page table walks (PTWs) 
that negatively impact application performance. Virtualized environments exacerbate \konreva{the} PTW latency \konreva{as they impose} two-level address translation (e.g., up to 24 memory accesses \konreva{can occur during a PTW} in a system with \konreva{n}ested \konreva{p}aging~\cite{amdnested, google-nested}), 
resulting in even higher address translation overhead\konreva{s} compared to native execution environments. Therefore, it is crucial to increase the \konreva{\emph{translation reach}} (i.e., the maximum amount of memory that can be covered by the \konreva{processor's TLB hierarchy}) 
to improve the effectiveness of TLBs and thus minimize PTWs. \konreva{Doing so} becomes increasingly important as PTW latency continues to rise with modern processors' deeper \konreva{multi-level page table} (PT) designs 
(e.g., 5-level radix PT in the latest Intel processors~\cite{5-levelpaging}).

Previous works have proposed various solutions to \konreva{reduce} the high cost of address translation and increase the translation reach of the \konreva{TLBs} such as \konreva{employing}
(i)  large hardware TLBs\VMtlblthree~\konreva{or} (ii) \konreva{backing up the last-level TLB with} a large software-managed TLB\VMsoftwareTLB. 
\konreva{Unfortunately, both solutions have significant drawbacks: increased access latency, power\konrevb{,} and area \konreva{(for hardware TLBs)}, and costly memory accesses, 
the need for large contiguous memory blocks, and complex OS modifications  \konreva{(for software-managed TLBs)}.}

\textbf{\konreva{Drawback of Large Hardware TLBs.}} 
 First, \konreva{a larger TLB has larger access latency} (\konreva{e.g.,} 1.4x larger latency for every 2x increase in size as reported by CACTI 7.0~\cite{cacti}), 
which may partially or entirely offset the performance gains \konreva{due to fewer} TLB misses. 
Second, a larger TLB leads to \konreva{larger chip area} and higher power consumption, resulting in higher costs and challenges in managing power constraints within the system. 
Third, \konreva{a larger TLB} may only benefit a specific subset of workloads, making it challenging to justify its applicability in a general-purpose system
\konreva{where some workloads are not sensitive to address translation performance}. \konrevb{Section~\ref{sec:motivation-htlb}} provide\konreva{s} a detailed quantitative analysis 
\konreva{of} (i) increasing the size \konreva{of a conventional} last-level TLB and (ii) expanding the TLB hierarchy with a hardware L3 TLB, \konreva{considering both realistic and optimistic (i.e, ideal) TLB designs}.

\textbf{\konreva{Drawbacks of Large Software-Managed TLBs.}}
 \konreva{First, to look up a software-managed TLB (STLB), the processor fetches STLB entries from the main memory into the cache hierarchy, resulting in a translation latency comparable to that of a PTW. 
Hence, \konrevf{an STLB} is more effective when PTW latency is higher than \konrevf{the} STLB access latency (e.g., \konrevf{as in} virtualized environments).}
Second, storing an STLB in memory requires allocating large contiguous memory blocks during runtime (\konrevb{on the} order of 10's of MB~\cite{pomtlbISCA2017}).
Third, an STLB introduces complex hardware/software interactions (e.g., evicting data from a hardware TLB to an STLB) and requires modifications in OS software. 
\konrevb{Section~\ref{sec:motivation-stlb}} provides a detailed quantitative analysis of STLBs.

\textbf{\textit{Opportunity}: Leveraging the Cache Hierarchy.} Rather than expanding hardware TLBs or introducing large software-managed TLBs, a cost-effective method to drastically increase translation reach is to 
 store \konreva{the} existing TLB entries within the existing cache hierarchy. For example, a 2MB L2 cache can fit \konreva{$128\times$} \konreva{the} TLB entries a \konreva{2048-entry} L2 TLB \konreva{holds}.  
 When a TLB entry resides inside the L2 cache, only one low-latency (\konrevb{e.g.,} $\approx$ 16 cycles) L2 access is needed to \konreva{find the} virtual-to-physical address translation instead of performing a high-latency 
 \konreva{(\konrevb{e.g.,}  $\approx 137$ cycles as shown in \S\ref{sec:motivation})} PTW. One potential pitfall of this approach is the potential reduction of caching capacity for application data, which could ultimately harm end-to-end performance. 
 However, as we show in \S\ref{sec:motivation} and as shown in multiple \konreva{prior} 
 works~\cite{catch,ferdman2012,jalili2023harvesting,damov2021,graspHPCA2020,droplet,manycoreSC2018,Bera2022HermesAL,linedistillation,spillreceive,adaptiveQureshiISCA2007}, modern data-intensive workloads, tend to \konreva{(greatly)} underutilize the cache hierarchy, \konreva{especially the large L2/L3/L4 caches}. 
 This is \konreva{because} \konrevb{many} modern working sets exceed the capacity of the cache hierarchy and \konrevb{many} data accesses exhibit \konreva{low} spatial and temporal locality~\cite{damov2021,graspHPCA2020,Tesseract,droplet,NowatzkiISCA19,dynamicgran2012matan,manycoreSC2018}. 
 Therefore, the underutilized cache blocks can \konrevb{likely} be repurposed to store TLB entries without replacing \konreva{useful} program data and harming end-to-end application performance.

\textbf{Our goal} in this work is to increase the translation reach of the processor's \konrevb{TLB hierarchy} by leveraging the underutilized resources \konrevb{in} the cache hierarchy.
\konreva{We aim to design such a practical technique that}: (i) \konreva{is} effective in both native and virtualized execution environments, (ii) \konreva{does not require or rely on contiguous} physical allocations, (iii) \konreva{is} 
transparent to \konrevb{both application and OS software} and (iv) \konreva{has} low area, power, and energy \konreva{costs}.

To this end, we present \system, a new \textit{software-transparent} mechanism that drastically increases the translation reach of the \konreva{TLB} by leveraging the underutilized resources of the cache hierarchy. 
The \textbf{key idea} of \system is to repurpose L2 cache blocks to store clusters of TLB entries. \konreva{Doing so} provides an additional low-latency and high-capacity component to back up the last-level TLB and \konrevb{thus reduces PTWs}.
\system \konreva{has} two main components\konrev{.} First, a PTW cost predictor (PTW-CP) \konreva{identifies} costly-to-translate addresses based on the frequency and cost of the PTWs \konreva{they lead to}. 
\konrevb{Leveraging} the PTW-CP, Victima uses the valuable cache space only for TLB entries that correspond to costly-to-translate pages, reducing the impact on cached application data. Second, 
a TLB-aware cache replacement policy \konrev{prioritizes keeping TLB entries in the cache hierarchy} \konreva{by considering (i) the translation pressure (\konreva{e.g.}, high last-level TLB \konreva{miss rate}) and (ii) the reuse of the TLB entries.}
 
 \textbf{Key Mechanism.} \system gets triggered \konreva{on} last-level TLB misses and evictions. \konreva{On} a last-level TLB miss, if PTW-CP predicts that the page will be costly-to-translate in the future, 
 \system transforms the data cache block that contains the last-level PT entries (PTEs) (fetched during the PTW) 
 \konreva{into a cluster of TLB entries} to enable direct access to the \konreva{corresponding} PTEs using a virtual address without walking the PT. 
 \konrevb{On} a last-level TLB eviction, if PTW-CP makes a positive prediction, \konreva{\system issues a PTW in the background to bring the PTEs of the evicted address into the L2 cache, and \system transforms the fetched PTE entries into a TLB entry.}
 This way, if the evicted TLB entr\konreva{y} \konreva{is} accessed again in the future, \system can directly access the corresponding PTE without walking the PT.
 \system (i) is effective in both native and virtualized environments, (ii) is completely transparent to \konreva{application and system} software,
  (iii) \konrev{unlike large software-managed TLBs}, does not require contiguous physical allocations, and (iv) is \konreva{compatible} with modern large page mechanisms (e.g., Transparent Huge Pages in Linux~\cite{hugepage}).

\textbf{Key Results.} We evaluate \system with an extended version of the Sniper simulator~\cite{sniper} \konreva{(which is open-source~\cite{victima_ae})} using 11 data-intensive applications from five diverse benchmark suites 
\konreva{(}GraphBIG~\cite{Lifeng2015}, GUPS~\cite{Plimpton2006}, XSBench~\cite{Tramm2014}, DLRM~\cite{dlmr} and GenomicsBench~\cite{genomicsbench}\konreva{)}.
Our evaluation yields \konreva{four major} results that show \system's effectiveness. 
\konreva{First}, in native execution environments, \system improves performance by \speedupoverbaseline\% on average over the baseline system that uses a four-level radix-tree-based PT, 
yielding 3.3\% and \speedupoverpomtlb\% higher performance compared to a system with an optimistic 64K-entry L2 TLB and a system with a state-of-the-art software-managed L3 TLB~\cite{pomtlbISCA2017}, respectively. At the same time, 
\system delivers \konreva{similar} performance as a system that employs an optimistic 128K-entry L2 TLB, while avoiding the associated area and power overheads.
\konreva{Second}, in virtualized environments, \system improves performance by \speedupoverbaselinevirt\% over the baseline nested paging mechanism~\cite{amdnested}, and \konreva{ outperforms an ideal shadow paging mechanism~\cite{vm25} by \speedupoverispvirt\% 
and a system that employs a state-of-the-art software-managed TLB~\cite{pomtlbISCA2017} by \speedupoverpomtlbvirt\%}.
\konreva{Third}, \system achieves such performance benefits by reducing L2 TLB miss latency by \latencyoverbaselinenative\% (\latencyoverbaselinevirt\%) on average in native (virtualized) execution environments compared to the baseline system (nested paging~\cite{amdnested}). 
\konreva{Fourth}, all of \system's benefits come at a modest cost of \area\% area overhead and \power\% power overhead compared to a modern high-end CPU~\cite{raptor_lake}.

\label{sec:contrib}
This paper makes the following \konreva{major} contributions:
\begin{itemize}
\item We observe a \konreva{new} opportunity to reuse the existing underutilized cache resources in order to store TLB entries 
and \konreva{increase the translation reach of the processor's TLB hierarchy at low cost and low overheads}.
\item We propose \system, a new \textit{software-transparent} mechanism that drastically increases the translation reach of the processor by \konreva{carefully and practically} leveraging the underutilized resources of the cache hierarchy. 
The \textbf{key idea} of \system is to repurpose L2 cache blocks to store clusters of TLB entries \konreva{for costly-to-translate pages}, \konreva{thereby} providing an additional low-latency 
and high-capacity component to back up the last-level TLB and \konreva{reducing} the number of PTWs.
\item We evaluate \system using a diverse set of data-intensive applications and demonstrate its effectiveness in both native and virtualized environments. 
\system achieves high performance benefits by effectively reducing last-level TLB miss latency compared to \konreva{both realistic and optimistic baseline systems}, with \konreva{very modest} area and power overheads compared to a modern high-end CPU.
\item We open-source Victima and all necessary traces and scripts to \konreva{completely} reproduce results at \textcolor{blue}{\url{https://github.com/CMU-SAFARI/Victima}}.
\end{itemize}
\vfill
\pagebreak

\section{Background} \label{sec:background}

\subsection{The Virtual Memory Abstraction}
Virtual memory is a cornerstone of most modern computing systems
\konrevd{that} eases the programming model by providing a \konrevb{convenient} abstraction to \konrevd{manage} the physical memory\VMold. The operating system (OS), transparently \konrevb{to application software}, maps each virtual memory address to its corresponding physical memory address. \konrevb{Doing so} 
{provides a number of benefits, including: (i) application-transparent memory management, (ii) sharing data between applications, (iii) process isolation, and (iv) \konrevb{page-level} memory protection.}
Conventional virtual memory designs allow any virtual page to map to any free physical page. Such a flexible address mapping 
enables two important key features of virtual memory: (i) efficient memory utilization, and (ii) sharing pages between applications. However,
\konrevb{such a} flexible address mapping \konrevb{mechanism} has a critical downside: it creates the need to store a large number of virtual-to-physical mappings, \konrevb{as for every process, the OS needs to store the physical
location of every virtual page}.

\subsection{Page Table (PT)}

{The \konrevb{PT} is a per-process data structure that stores the mappings between virtual and physical pages. 
In modern x86-64 processors, the \konrevb{PT} is organized as a four-level radix-tree~\cite{intelx86manual}. Even though the radix-tree-based PT 
optimizes for storage efficiency, it requires \konrevb{multiple} pointer-chasing operations to discover the virtual-to-physical mapping.  
To search for a virtual-to-physical address mapping, the system needs to \emph{sequentially} access each of the four levels of the page table. 
This process is called \emph{page table walk (PTW)}.}

\begin{figure}[h]
    \vspace{-2mm}
    \centering
    \includegraphics[width=\columnwidth]{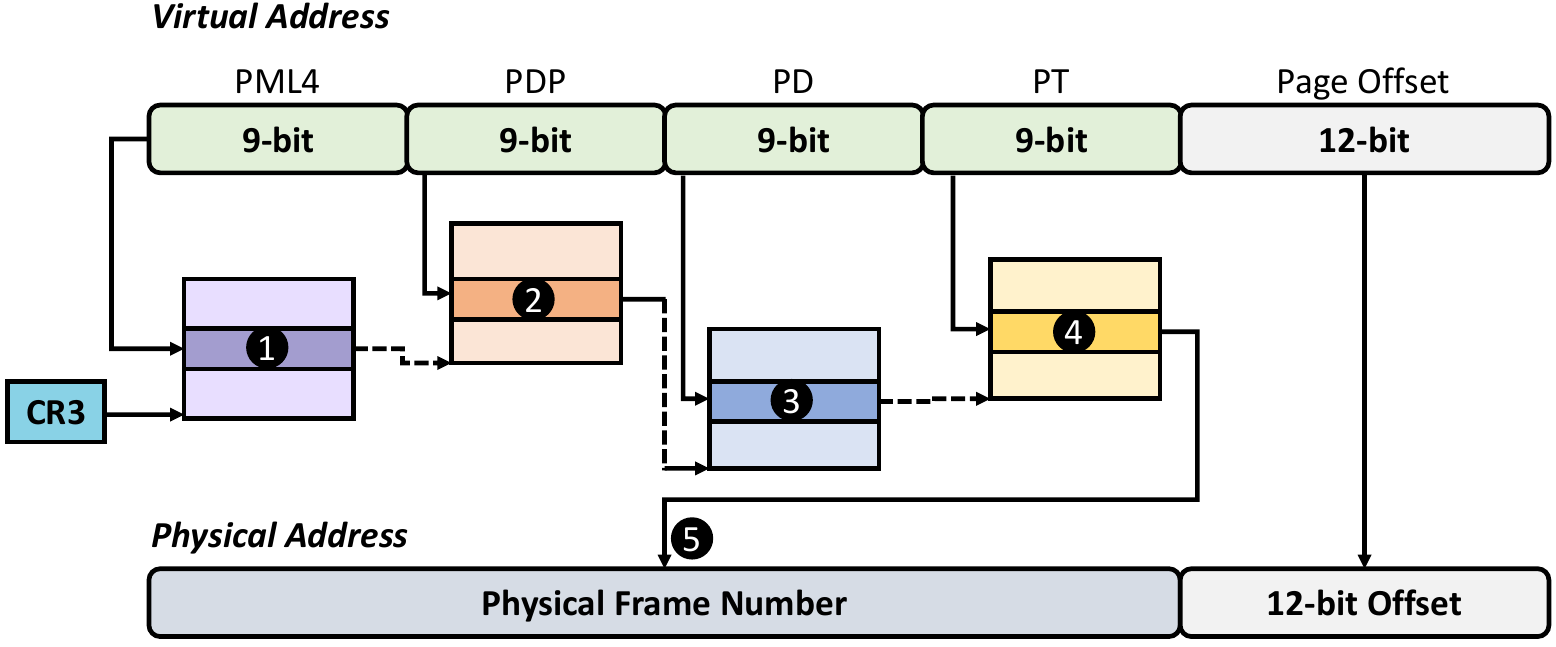}
    \vspace{-6mm}
    \caption{Four-level radix-tree page table walk in x86-64 ISA.}
    \label{fig:page-table}
    \vspace{-3mm}
\end{figure}

{Figure~\ref{fig:page-table} shows the PTW assuming (i) an x86-64 four-level radix-tree PT whose base address is stored in the CR3 register, and (ii) 4KB pages.
 As shown in Figure~\ref{fig:page-table}, a single PTW requires four sequential memory accesses \konrevb{\circled{1}-\circled{4}} to discover the physical page number. 
 The processor uses the first 9-bits of the virtual address as offset \konrevb{(Page Map Level4; PML4)} to index the appropriate entry of the PT within the first level of the PT~\circled{1}. 
 The processor then reads the pointer stored in the first level of the PT to access the second-level of the PT~\circled{2}. 
 It uses the next 9-bit set \konrevb{(Page Directory Page table; PDP)} from the virtual address to locate the appropriate entry within the second level. 
 This process continues \konrevb{iteratively} for each subsequent level of the PT \konrevb{(Page Directory; PD~\circled{3} and Page Table; PT~\circled{4})}.
Eventually, the processor reaches the leaf level of the PT, where it finds the final entry containing the physical page number corresponding to the given virtual address~\circled{5}. 
ARM processors use a similar approach, with the number of levels varying across different versions of the ISA~\cite{arm-manual-tlbmaintenance}.}

\subsection{Virtualized Environments}

In virtualized environments, each memory request requires a two-level address translation: (i) from guest-virtual to guest-physical, and (ii) from guest-physical to host-physical. The dominant technique to perform address translation in virtualized environments is Nested Paging (NP)~\cite{amdnested, google-nested}. 
In NP, the system uses two page tables: the guest page table that stores guest-virtual to guest-physical address mappings and the host page table that stores guest-physical to host-physical address mappings.  
To search for the mapping between a guest-virtual page to a host-physical page, NP performs a two-dimensional walk, {since} a host page table walk is required for each level of the guest page table walk.  
Therefore, in a virtualized environment with a four-level radix-tree-based PT, NP-based address translation can cause up to 24 sequential memory accesses (a 6$\times$ increase in memory accesses compared to the native execution environment).

\subsection{Memory \konrevb{Management} Unit (MMU)}
\begin{figure}[h]
    \vspace{-2mm}

    \centering
    \includegraphics[width=3.3in]{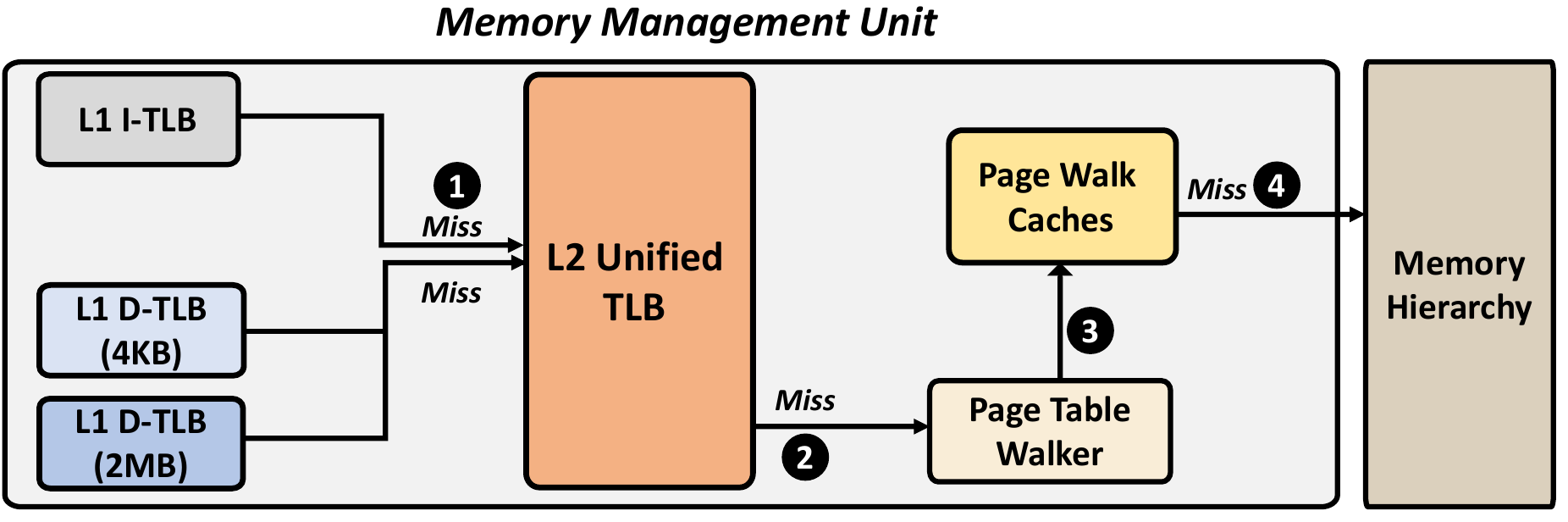}
    \vspace{-2mm}
    \caption{Structure of the Memory Management Unit (MMU) of a modern processor.}
    \label{fig:mmu}
    \vspace{-3mm}

\end{figure}

\konrevb{When} a user process {generates} a memory (i.e., instruction or data) request, the processor needs to translate the virtual address to its corresponding physical address. 
Address translation is a critical operation because it sits on the critical path of the memory access flow: no memory access is possible unless the requested virtual address is first translated into its corresponding physical address.
Given that frequent PTWs lead to high address translation overheads, modern cores comprise of a specialized memory management unit (MMU) responsible for accelerating address translation.
\autoref{fig:mmu} {shows an example structure of the MMU of a modern processor~\cite{cascadelake}, consisting of three key components: (i) a two-level hierarchy of translation lookaside buffers (TLBs), 
(ii) a hardware page table walker, and (iii) page walk caches (PWCs).}

L1 TLBs are {highly- or fully-}associative caches that directly provide the physical address for recently-accessed virtual pages at very low latency (i.e., typically within 1 cycle). There are two separate L1 TLBs, one for instructions (L1 I-TLB) and one for data (L1 D-TLB). 
{Modern TLBs make use of multiple page sizes beyond 4KB in order to (i) cover large amounts memory with a single entry and (ii)  maintain compatibility with modern OSes that transparently allocate large pages~\cite{tridentMICRO2021,corbet2011,reserve,panwar2019hawkeye}. 
For example, \konrevb{an Intel} Cascade Lake core~\cite{cascadelake} employs 2 L1 D-TLBs, \konrevd{one for 2MB pages and one for 4KB pages.}}
Translation requests that miss in the L1 TLBs~\circled{1} are forwarded to a unified L2 TLB, that stores translations for both instructions and data. 
In case of an L2 TLB miss, the MMU triggers a PTW~\circled{2}. 
PTW is performed by a dedicated hardware page table walker capable of performing multiple concurrent PTWs.
In order to reduce PTW latency, page table walkers are equipped with page walk caches (PWC)~\circled{3}, which are small dedicated caches for each level of the PT (for the first three levels in x86-64). 
In case of a PWC miss, the MMU issues the request(s) for the corresponding level of the PT to the conventional memory hierarchy~\circled{4}.

\begin{figure}[h]
    \vspace{-2mm}

    \centering
    \includegraphics[width=3.3in]{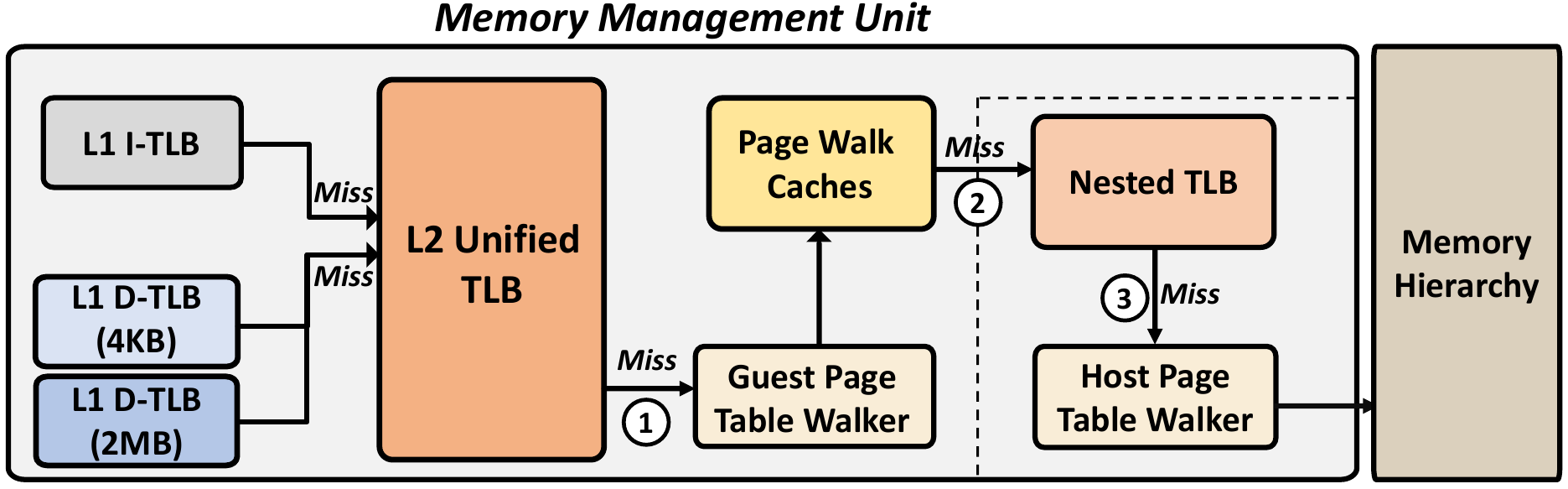}
    \vspace{-2mm}

    \caption{MMU extensions to support address translation in virtualized environment\konrevb{s} using Nested Paging~\cite{amdnested}.}
    \label{fig:mmu_virt}
    \vspace{-3mm}

\end{figure}

To accelerate address translation in virtualized execution environments \konrevb{that use} Nested Paging~\cite{amdnested}, {as shown in Figure~\ref{fig:mmu_virt}}, the MMU is additionally equipped with (i) a nested TLB that stores guest-physical-to-host-physical mappings 
and (ii) an additional hardware page table walker that walks the host PT (while the other one walks the guest PT).
{Upon an L2 TLB miss, the MMU  triggers a guest PTW to retrieve the guest-physical address~\circledwhite{1}.
\konrevb{On a PWC miss, the guest Page Table Walker must retrieve the guest PT entries from the cache hierarchy. 
However, to access the cache hierarchy that operates on host-physical addresses, the guest PTW must first translate the host-virtual address to the host-physical address using a host PTW.}
To avoid the host PTW, the MMU probes the nested TLB to search for the host-virtual-to-host-physical translation~\circledwhite{2}.
Only in case of a nested TLB miss the MMU triggers the host PTW~\circledwhite{3}.}

\section{Motivation}
\label{sec:motivation}
\begin{figure}[h]
    \vspace{-3mm}
    \centering
    \includegraphics[width=\linewidth]{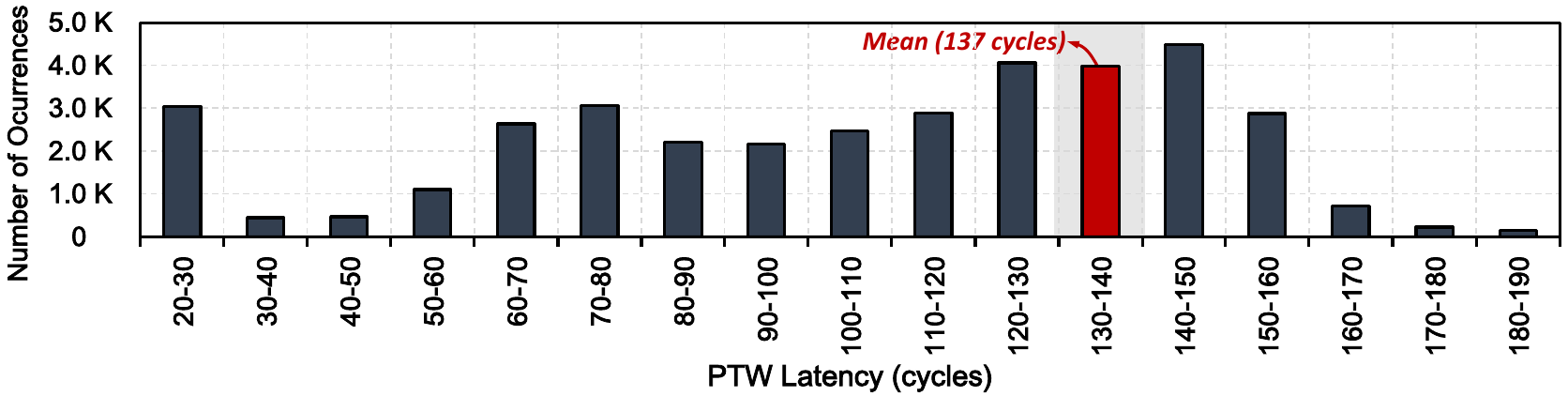}
    \vspace{-7mm}
    \caption{Distribution of PTW latency.}
    \label{fig:ptw_latency}
    \vspace{-3mm}
\end{figure}

As shown in multiple prior academic works and industrial studies\VMcharacterization, \konrevc{various} modern data-intensive workloads experience severe performance bottlenecks due to address translation. 
\konreve{For example, a system that (i) employs a 1.5K-entry L2 TLB and (ii) uses both 4KB and 2MB pages,} experiences \konrevc{a} high MPKI \konrevc{of} 39, \konrevc{averaged} across all \konrevc{evaluated} workloads (see Fig.~\ref{fig:l2-tlb-mpki}).\footnote{\S\ref{sec:methodology} describes our evaluation methodology in detail.} 
At the same \konrevc{time}, as we show in Figure~\ref{fig:ptw_latency},  the average latency of a PTW is 137 cycles.\footnote{The x-axis of Figure~\ref{fig:ptw_latency} is cut off (at 190 cycles) since only 
0.2\% of the PTWs take more than 190 cycles to complete. Maximum observed PTW latency is 608 cycles.}
\konreve{Based on our evaluation results, frequent L2 TLB misses in combination with high-latency PTWs lead to an average of 30\% of total execution cycles spent on address translation.}

Previous works \konrevc{propose} various solutions to {reduce} the high cost of address translation and increase the translation reach of the {TLBs} such as {employing}
(i) large hardware TLBs\VMtlblthree~or (ii) backing up the last-level TLB with a large software-managed TLB\VMsoftwareTLB. 
\konrevc{We examine these solutions and their shortcomings in \S\ref{sec:motivation-htlb} and \S\ref{sec:motivation-stlb}.}

\subsection{Large Hardware TLBs} 
\label{sec:motivation-htlb}
\begin{figure}[h]
    \vspace{-2mm}
    \centering
    \includegraphics[width=\linewidth]{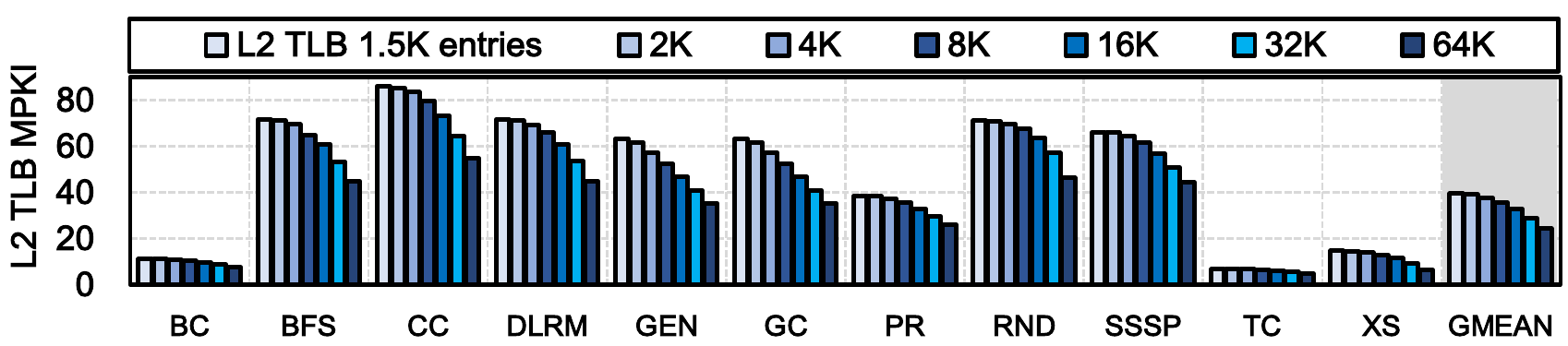}
    \vspace{-8mm}
    \caption{L2 TLB MPKI \konrevc{for} L2 TLBs with \konrevc{different} sizes.}
    \label{fig:l2-tlb-mpki}
    \vspace{-2mm}
\end{figure}

\begin{figure}[h]
    \centering
    \includegraphics[width=\linewidth]{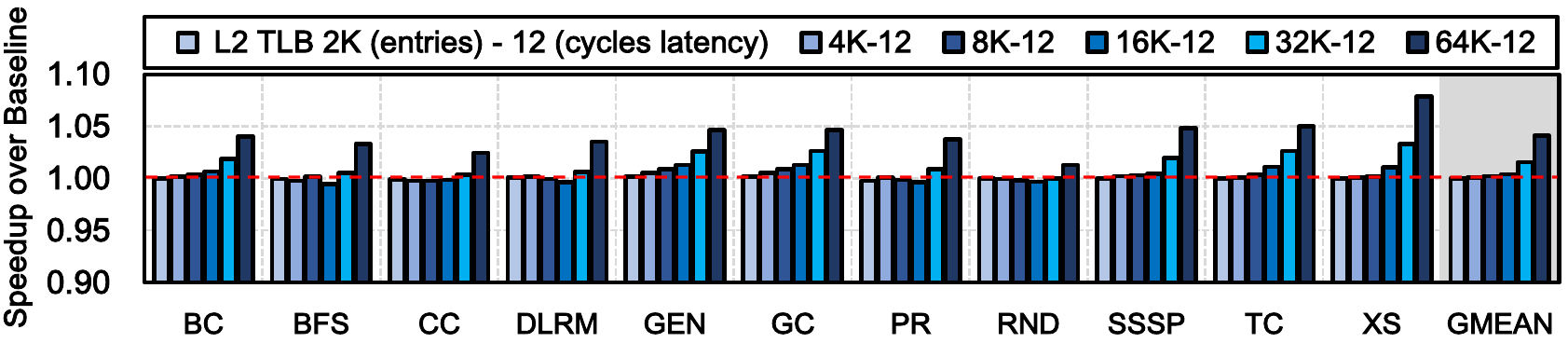}
    \vspace{-7mm}
    \caption{\konrevc{Speedup provided by larger L2 TLBs with equal access latencies (i.e., 12 cycles) over the baseline system (1.5K-entry L2 TLB).}}
    \label{fig:l2-tlb-perf-opt}
    \vspace{-4mm}
\end{figure}

We evaluate the effectiveness of increasing the size of the TLB. Our methodology and workloads are described in detail in \S\ref{sec:methodology}.
\autoref{fig:l2-tlb-mpki} demonstrates the L2 TLB MPKI as we increase the size of the L2 TLB from 1.5K up to 64K entries. 
\konrevc{We observe that} increasing the number of L2 TLB entries from 1.5K to 64K \konrevc{(i.e., by 42$\times$)} 
results in reducing the MPKI from 39 to 24 \konrevc{(i.e., by 44\%)}.
To better understand the potential performance of increasing the size of the L2 TLB, Figure \ref{fig:l2-tlb-perf-opt}
shows the \konrev{execution time} speedup of L2 TLB configurations with increasing sizes but equal access latencies (i.e., 12 cycles) compared to the baseline system (1.5K-entry L2 TLB). 
\konrevc{We evaluate an optimistic setting where the access latency is set to 12 cycles \emph{regardless} of the TLB size.}
\konrevc{We observe that} the optimistic 64K-entry configuration (that reduces MPKI by 44\%) 
leads to a 4.0\% higher performance \konrevc{on average} compared to the baseline \konrev{1.5K-entry L2 TLB} configuration.

\begin{figure}[h]
\vspace{-2mm}
\centering
\includegraphics[width=\linewidth]{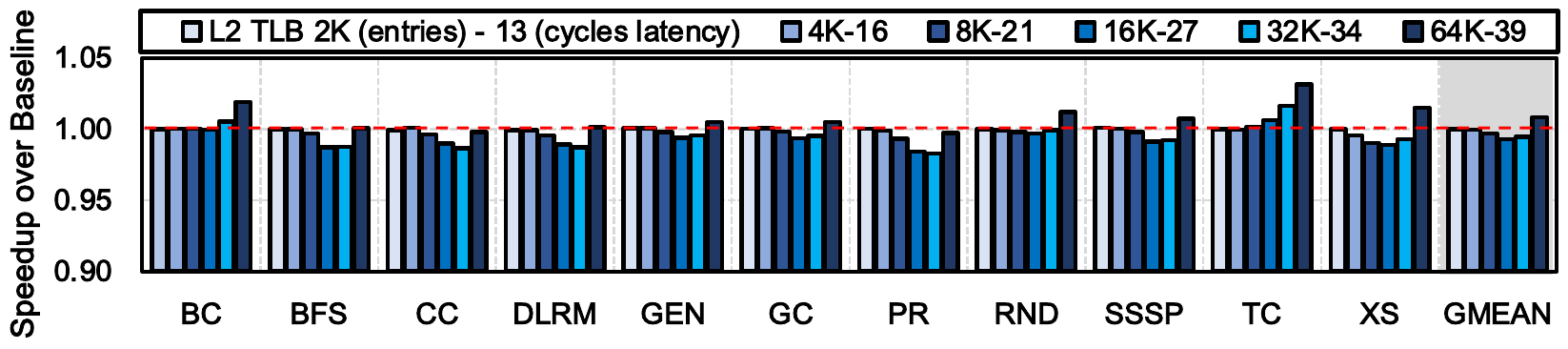}
\vspace{-6mm}
\caption{\konrevc{Speedup provided by larger L2 TLBs over the baseline system (1.5K-entry L2 TLB). L2 TLB access latency is adjusted based on the size of the TLB (modeled using CACTI 7.0~\cite{cacti}).}}
\label{fig:l2-tlb-perf-real}
\vspace{-2mm}
\end{figure}

\konrevc{Unfortunately,} increasing the TLB size does not come for free: \konrevc{it} leads to \konrevc{larger access} latency \konrevc{(as well as area and power)}, 
which counteracts the potential performance benefits due to fewer PTWs. For instance, according to CACTI 7.0~\cite{cacti}, the latency of accessing a \konrevc{64K-entry} large TLB is as high as 39 cycles. 
Figure~\ref{fig:l2-tlb-perf-real} shows the \konrevc{execution time} speedup of realistic L2 TLB configurations with increasing sizes, while the access latency is adjusted based on the size of the TLB 
\konrevc{(based on CACTI 7.0 modeling~\cite{cacti})}, compared to the baseline system (1.5K-entry L2 TLB \konrevc{with 12-cycle access latency}).
\konrevc{We observe that} in this \konrevc{realistic setting}, the performance benefits of increasing the L2 TLB size are significantly lower compared to the optimistic \konrevc{setting (Fig.~\ref{fig:l2-tlb-perf-opt})}. 
The realistic 64K-entry configuration (that reduces MPKI by 44\%, \konrevc{but comes with a 39-cycle access latency})  leads to \konrevc{only} 0.8\% higher \konrevc{average} performance \konrevc{over} the baseline configuration.
\konrevc{We conclude that although increasing the L2 TLB size reduces PTWs, it comes with increased access latency (as well as power and area), which leads to small performance benefits realistically. }

\begin{figure}[h]
\centering
\includegraphics[width=\linewidth]{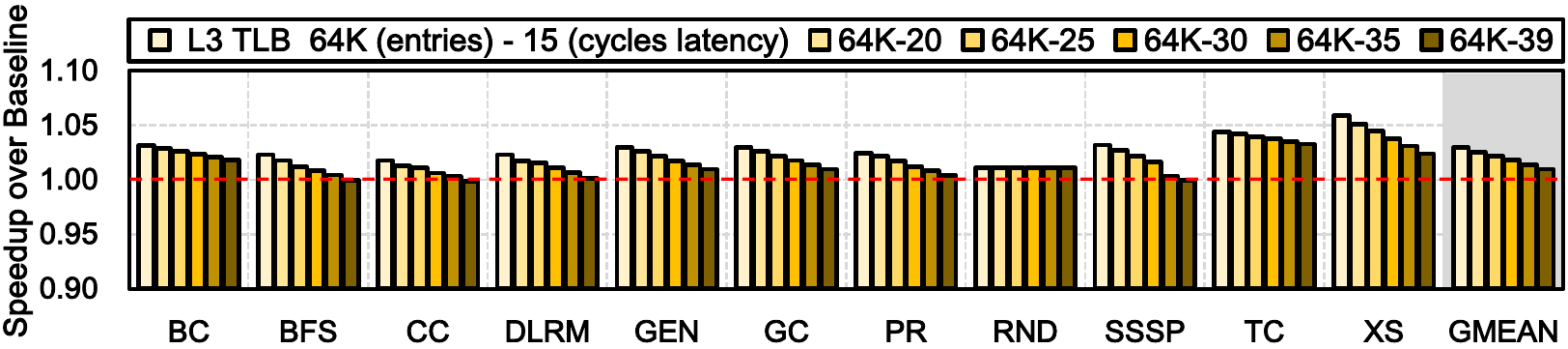}
\vspace{-6mm}
\caption{\konrevc{Speedup provided by adding a 64K-entry L3 TLB with different access latencies over the baseline system.}}
\label{fig:l3tlb}
\vspace{-4mm}

\end{figure}

Increasing the size of the L2 TLB has a negative impact on the translation latency of 
requests that hit in the L2 TLB. Therefore, to keep the access latency of the L2 TLB \konrevc{small}, we also explore a scenario where the TLB hierarchy is extended with a large hardware L3 TLB. 
Figure~\ref{fig:l3tlb} shows the \konrevc{execution time} speedup achieved by a system with a 64K-entry L3 TLB with increasing access latencies, 
ranging from 15 cycles up to 39 cycles (\konrevc{which is the} latency suggested by CACTI 7.0~\cite{cacti}), 
compared to the baseline system that employs a two-level TLB hierarchy (with a 1.5K-entry 12-cycle L2 TLB). 
\konrevc{We observe that} a large 64K-entry L3 TLB with \konrevc{a very aggressive} 15-cycle access latency leads to a 2.9\% performance increase compared to the baseline system.
The performance gains are lower compared to employing a 64K-entry L2 TLB (4.0\%). This is because, for applications that experience low L2 TLB hit rates,
employing an L3 TLB results in \konrevc{a} higher L3 TLB hit latency (L2 TLB miss latency $+$ L3 TLB hit latency) compared to using a large L2 TLB. 
We conclude that employing a large L3 TLB is not universally beneficial, and the performance gains heavily \konrevd{depend} on the L2 TLB hit rates \konrevc{and L3 TLB access latencies}.

\subsection{Large Software-Managed TLBs}
\label{sec:motivation-stlb}

\begin{figure}[h]
\vspace{-2mm}
\centering
\includegraphics[width=\linewidth]{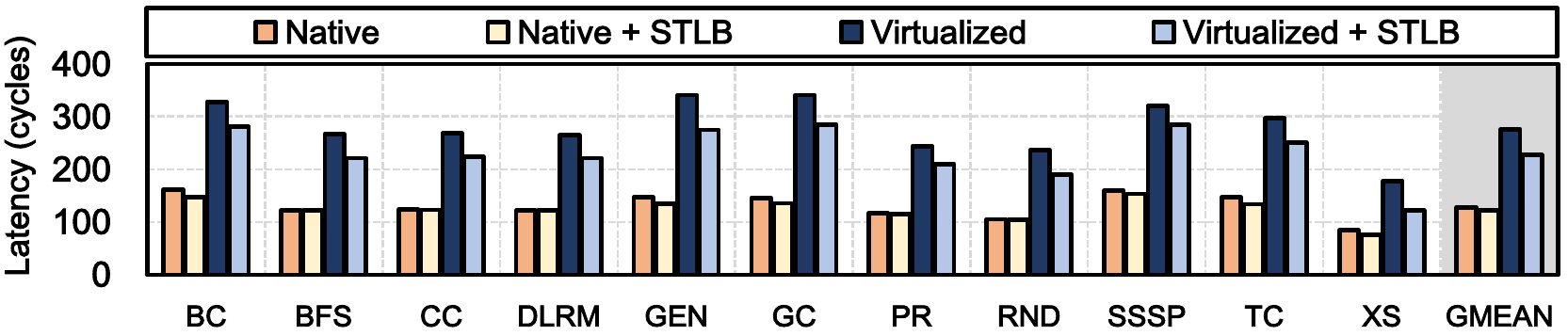}
\vspace{-6mm}
\caption{\konkanelloreva{L2 TLB miss latency of (i) baseline system in native execution, (ii) system with STLB~\cite{pomtlbISCA2017} in native execution,  (iii) baseline system in virtualized execution and, \konrevc{(iv) STLB in virtualized execution.} }}
\label{fig:stlb}
\vspace{-2mm}
\end{figure}

Previous works\VMsoftwareTLB~propose using large software-managed TLBs to reduce PTWs. 
However, software-managed TLBs suffer from four key disadvantages. 
First, to look up a software-managed TLB (STLB), the processor fetches STLB entries from the main memory into the cache hierarchy. 
At the same time, the hit rate of STLBs \konrevc{likely} does not justify the cost of fetching STLB entries from the main memory. 
Hence, the total latency of accessing STLB entries and performing PTWs is comparable to the latency of performing PTWs in the baseline system.
To validate our claim, Figure~\ref{fig:stlb} shows the average L2 TLB miss latency in (i) the baseline system in native execution, 
(ii) a system with a state-of-the-art L3 STLB~\cite{pomtlbISCA2017} in native execution, (iii) the baseline system that employs nested paging (NP)~\cite{amdnested} in virtualized execution
and \konrevc{(iv) a system with a state-of-the-art L3 STLB~\cite{pomtlbISCA2017} and NP~\cite{amdnested} in virtualized execution.}
\konrevd{We observe} that the average L2 TLB miss latency in a system with an STLB is 122 cycles, which is comparable to the baseline system (128 cycles).
However, the average L2 TLB miss latency in the system with NP in virtualized execution is 275 cycles, which is higher than the average L2 TLB miss latency in a system with an L3 STLB (220 cycles) in virtualized execution, making 
the STLB a more attractive solution in virtualized execution environments.

Second, allocating an STLB in software requires contiguous physical address space (\konrevd{on} the order of 10's of MB), which is \konrevc{difficult} to find in environments where 
memory is heavily fragmented, such as data centers~\cite{contiguitas2023,translationranger2019,chloe2020} \konrevc{and in cases where memory capacity pressure is high~\cite{tmoASPLOS2022,memtrade2023sigmetrics,softwaredefinedASPLOS2019}}. 
Third, resizing an STLB throughout the execution of the program to match the program's needs is challenging due to the large data movement 
cost of migrating the TLB entries betweeen different software data structures~\cite{elastic-cuckoo-asplos20,mehtJovanHPCA2023,flataAsplos2022,nestedecht}. 
Fourth, integrating a software-managed TLB in the address translation pipeline requires OS and hardware changes to support 
(i) flushing and updating software STLB entries during a TLB shootdown~\cite{csaltMICRO2017,pomtlbISCA2017}, (ii) handling evictions from the hardware TLB to the STLB~\cite{csaltMICRO2017,pomtlbISCA2017}.

\subsection{Opportunity: Storing TLB Entries Inside the Cache Hierarchy}
    
\konrevc{Instead of} expanding hardware TLBs or introducing large software-managed TLBs, \konrevc{we posit that} a cost-effective method to drastically \konrevc{increase the translation reach of the TLB hierarchy} is to 
store {the} existing TLB entries within the existing cache hierarchy. For example, a 2MB L2 cache can fit {$128\times$} {the} TLB entries a {2048-entry} L2 TLB {holds}.  
When a TLB entry resides inside the L2 cache, only one low-latency (i.e., $\approx$ 16 cycles) L2 access is needed to {find the} virtual-to-physical address translation instead of performing a high-latency 
{(i.e., $\approx 137$ cycles on average)} PTW. 
    
\begin{figure}[h]
\vspace{-2mm}
\centering
\includegraphics[width=\linewidth]{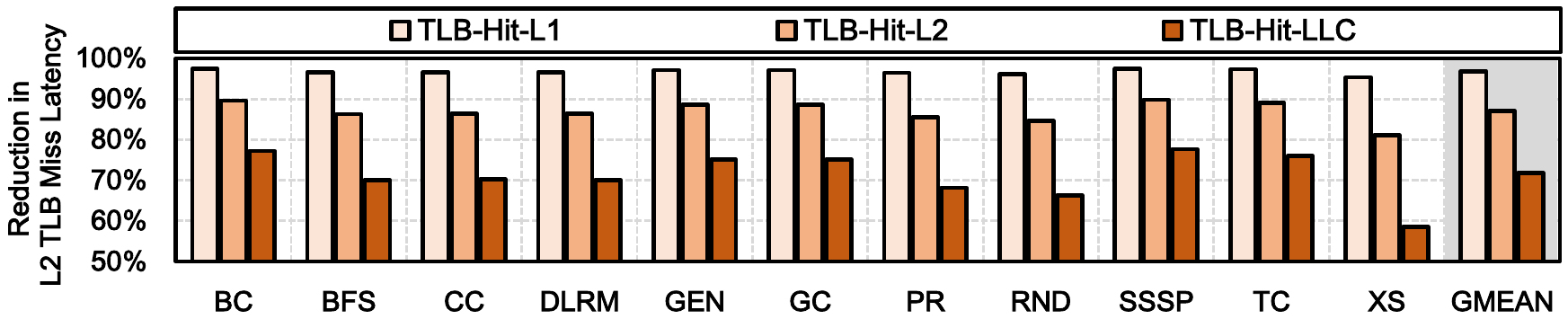}
\vspace{-7mm}
\caption{\konrevc{Reduction in L2 TLB miss latency when L1/L2/LLC serve all L2 TLB misses over the baseline system.}}
\label{fig:victima_ideal}
\vspace{-3mm}
\end{figure}

To better understand the potential of caching TLB entries in the cache hierarchy, we conduct a study where for every L2 TLB miss, the translation request is \emph{always} served from the 
L1 cache (\emph{TLB-hit-L1}), L2 cache (\emph{TLB-hit-L2}) or the LLC (\emph{TLB-hit-LLC}). \autoref{fig:victima_ideal} shows the reduction \konrevc{in address} translation latency \konrevc{provided by} TLB-hit-\{L1, L2, LLC\} 
compared to the baseline system. \konrevc{We observe that}, even when servicing every L2 TLB miss from the LLC (\nb{which} takes $\approx$35 cycles to access), \konrevc{L2 TLB miss latency} is reduced by 
71.9\% on average across 11 workloads. \konrevd{We conclude} that caching TLB entries inside the cache hierarchy \konrevc{can} \konrevc{potentially greatly reduce the} address translation latency. 
 
\subsection{Cache Underutilization}
One potential pitfall of \konrevc{storing TLB entries inside the cache hierarchy} is the potential reduction of caching capacity for application data, which could ultimately harm end-to-end performance. 
However, as shown in {prior} works~\cite{catch,ferdman2012,jalili2023harvesting,damov2021,graspHPCA2020,droplet,manycoreSC2018,Bera2022HermesAL,linedistillation,spillreceive,adaptiveQureshiISCA2007},
\konrevc{many} modern data-intensive workloads, tend to {(greatly)} underutilize the cache hierarchy, {especially \konrevc{the} large L2/L3/L4 caches}. 
This is {because} modern working sets exceed the capacity of the cache hierarchy and data accesses exhibit {low} spatial and temporal locality~\cite{damov2021,graspHPCA2020,Tesseract,droplet,NowatzkiISCA19,dynamicgran2012matan,manycoreSC2018}. 

\begin{figure}[h]
\vspace{-2mm}
\centering
\includegraphics[width=\linewidth]{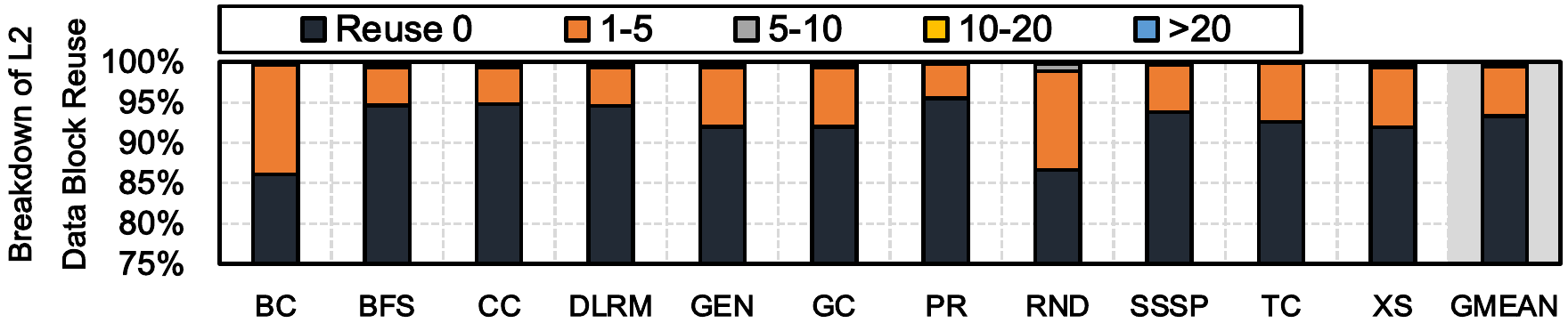}
\vspace{-7mm}
\caption{\konrevc{Reuse-level distribution of L2 cache blocks.}}
\label{fig:l2-reuse}
\vspace{-3mm}
\end{figure}

\autoref{fig:l2-reuse} shows the reuse-level \konrevd{distribution} of blocks in the L2 cache across \konrevc{our evaluated} \nb{data-intensive} workloads \konrevc{(note that y-axis starts from 75\%)}.
We observe that on average 92\% of the cache blocks experience \konrevc{no} reuse \konrevc{(i.e., 0 reuse)} after being brought to the L2 cache (i.e., \konreve{these blocks are \emph{not}} accessed while they reside inside the L2 cache). \konrevd{In contrast,} only 
8\% of blocks experience reuse higher than 1 (i.e., \konrevd{they are accessed} more than once while they reside inside the L2 cache).
We conclude that a large \konrevc{fraction of the} underutilized cache blocks can be repurposed to store TLB entries \emph{without} replacing {useful} program data and harming end-to-end application performance.

\subsection{Our Goal}
\textbf{Our goal} is to increase the translation reach of the processor's \konrevb{TLB hierarchy} by leveraging the underutilized resources in the cache hierarchy.
\konreva{We aim to design such a practical technique that}: (i) \konreva{is} effective in both native and virtualized execution environments, (ii) \konreva{does not require or rely on contiguous} physical allocations, (iii) \konreva{is} 
transparent to \konrevb{both application and OS software} and (iv) \konreva{has} low area, power, and energy \konreva{costs}. \konrevc{To this end, our key idea is to store TLB entries in the cache hierarchy.}

\vspace{-2mm}
\section{\system: Design Overview}


%

\konrevc{We present \system}, a new \textit{software-transparent} mechanism that drastically increases the translation reach of the \konreva{TLB} by leveraging the underutilized resources of the cache hierarchy.
The \textbf{key idea} of \system is to repurpose L2 cache blocks to store clusters of TLB entries. \konreva{Doing so} provides an additional low-latency and high-capacity component to back up the last-level TLB and \konreva{thus reduces PTWs}. 
\system \konreva{has} two main components\konrev{.} First, a PTW cost predictor (PTW-CP) \konreva{identifies} costly-to-translate addresses based on the frequency and cost of the PTWs \konreva{they lead to}. 
\konreva{Leveraging} the PTW-CP, Victima uses the valuable cache space only for TLB entries that correspond to costly-to-translate pages, reducing the impact on cached application data. Second, 
a TLB-aware cache replacement policy \konrev{prioritizes keeping TLB entries in the cache hierarchy} \konrevc{by taking into account (i) the translation pressure (\konreva{e.g.}, high last-level TLB \konreva{miss rate}) and (ii) the reuse characteristics of the TLB entries.}

\begin{figure}[h!]
    \vspace{-3mm}
    \centering
    \includegraphics[width=\linewidth]{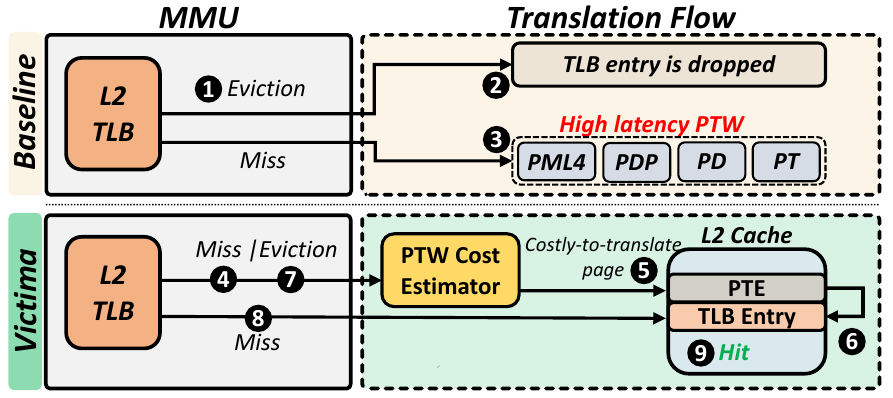}
    \vspace{-6mm}
    \caption{Address translation flow in a conventional baseline processor~\cite{raptor_lake} and \system.}
    \label{fig:overview}
    \vspace{-2mm}

\end{figure}

\autoref{fig:overview} shows the translation flow in \system compared to the one in a conventional baseline processor~\cite{raptor_lake}. 
In the baseline system \konrevc{(Fig.~\ref{fig:overview} top)} , (i) whenever an entry is evicted from the L2 TLB~\circled{1}, the evicted TLB entry is not cached anywhere. Hence, (i) the TLB entry is dropped~\circled{2} and (ii) a high-latency PTW is required to fetch it when it is requested again~\circled{3}. 
In contrast, \system \konrevc{(Fig.~\ref{fig:overview} bottom)} stores into the L2 cache (i) entries that get evicted from the L2 TLB and (ii) the TLB entries of memory accesses that cause L2 TLB misses. 
\system gets triggered \konreva{on} last-level TLB misses and evictions~\circled{4}. \konreva{On} a last-level TLB miss, if PTW-CP predicts that the page will be costly-to-translate in the future~\circled{5}, 
\system transforms the data cache block that contains the last-level PT entries (PTEs) (fetched during the PTW) 
\konreva{into a cluster of TLB entries}~\circled{6} to enable direct access to the \konreva{corresponding cluster of} PTEs using a virtual address without walking the PT.
Storing a cluster of TLB entries for contiguous virtual pages inside the L2 cache can be highly beneficial for applications whose memory accesses exhibit high spatial locality.
\konreva{On} a last-level TLB eviction~\circled{7}, if PTW-CP makes a positive prediction, \konreva{\system issues a PTW in the background to bring the PTEs of the evicted address into the L2 cache, and \system transforms the fetched PTEs into a TLB entry.}
This way, if the evicted TLB entr\konreva{y} \konreva{is} accessed again in the future~\circled{8}, \system can directly access the corresponding PTE from the \konrevd{L2 cache} without walking the PT~\circled{9}. 
 Storing evicted TLB entries in the L2 cache can be highly beneficial for applications that experience
a high number of capacity misses in the TLB hierarchy. \system's functionality \konrevc{seamlessly} applies to virtualized environments \konrevc{as well}. 
\konrevc{In virtualized execution,} where \system stores into the L2 cache both (i) conventional TLB entries that store direct guest-virtual-to-host-physical mappings 
as well as (ii) nested TLB entries that store guest-physical-to-host-physical mappings.

\section{\system: Detailed Design}
\label{sec:implementation}

We \konrevc{describe} in detail (i) how the L2 cache is modified to store TLB entries, 
(ii) how \system inserts TLB entries \konrevc{into} the L2 cache, 
(iii) how \konrevc{address} translation flow changes in the presence of \system, 
(iv) how \system operates in virtualized environments and 
(v) how \system maintains TLB entries coherent. 
\konrevc{We use as the reference design point a modern x86-64 system that employs 48-bit virtual addresses (VA) and 52-bit physical adddresses (PA)~\cite{intelx86manual}}.

\subsection{Modifications \konrevc{to the} L2 Cache}
\label{sec:l2-changes}
\konrevc{We minimally modify the L2 cache} to (i) support storing TLB entries and (ii) enable a TLB-aware replacement policy 
that favors keeping TLB entries inside the L2 cache taking into account  address translation pressure (e.g., L2 TLB MPKI) and the reuse characteristics of TLB entries. 

\head{\konrevd{TLB Blocks}}
\konrevd{We introduce a new cache block type to store TLB entries in the data store of the L2 cache, called the TLB block.}
\autoref{fig:layout} shows how the same address maps to (i) a conventional L2 data cache block and 
(ii) an L2 cache block that contains TLB entries for 4KB or 2MB pages. 
\konrevc{Each cache entry can potentially store a data block or a TLB block.}
\konrevc{A conventional data block is (typically)} accessed using the PA while \konrevc{a TLB block is} accessed using the VA. 
\system modifies the cache block metadata layout to enable storing TLB entries. 
{First, an additional bit is needed to distinguish between \konrevd{a data block versus a TLB block}. 
Second, in \konrevd{a} conventional data block, the size of the tag of a 1MB, 16-way associative L2 cache consists of $52-log_2(1024)-log_2(64)=36$ bits. 
However, in a TLB block, the tag consists of \konrevc{only} 23 bits and is computed as $48-log_2(4KB)-log_2(1024)-log_2(8)=23$ bits
which is smaller than the tag needed \konrevd{for a data block}.\footnote{\konrevc{Each 64-byte TLB block can store up to 8 8-byte PTEs. Victima uses the 3 least significant bits of the virtual page number to \konrevd{identify and access} a specific PTE.}} 
We leverage the unused space \konrevc{in TLB blocks} to (i) avoid \konrevc{aliasing} and (ii) store page size information.

\begin{figure}[h!]
    \vspace{-2mm}
    \centering
    \includegraphics[width=\linewidth]{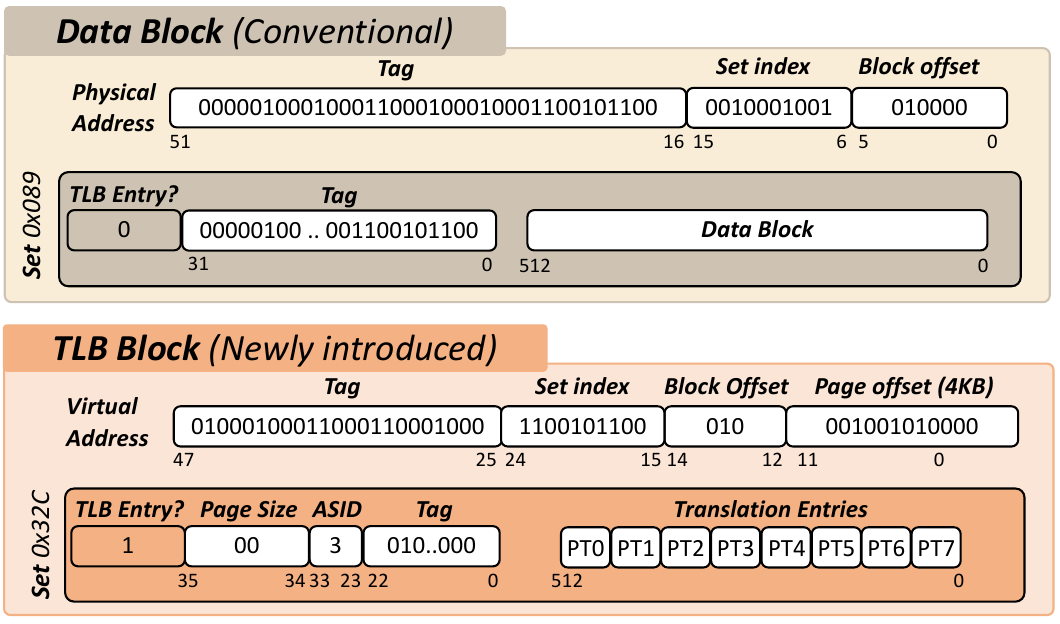}
    \vspace{-5mm}
    \caption{\konkanelloreve{\konrevc{Conventional data block layout} (top) and \konrevc{conventional} TLB block layout for the same address (bottom).}}
    \label{fig:layout}
    \vspace{-2mm}
\end{figure}

To prevent aliasing between the virtual addresses (VAs) of different processes, 
11 unused bits of the tag are reserved for storing the address-space identifier (ASID) or the virtual-machine identifier (VMID) of each process.
The rest of the bits are used \konrevc{to store} page size information.} 
{Given a 48-bit VA and 52-bit PA, we can spare 11 bits for the ASID/VMID. 
As the VA size becomes larger, e.g., 57 bits, fewer bits can be spared for the ASID/VMID (4 bits in case of 57-bit VA and 52-bit PA). 
However, modern operating systems do \emph{not} use more than 12 ASIDs/core~\cite{linuxasid} in order to avoid expensive lookups in the ASID table. 
Hence, when using 57-bit VAs and 52-bit PAs, even with \konrevc{only} 4 bits \konrevc{left} for the ASID, there is no risk of aliasing.}

\konrevc{For a cache with 64-byte cache lines, it is possible to uniquely tag and avoid aliasing between TLB entries \konrevd{(without increasing the size of the cache's hardware tag entries)} only if
$(PA_{length} > VA_{length}-9)$}.\footnote{\konrevd{If $PA_{length} \leq (VA_{length}-9)$, a single VA can map to different TLB blocks. 
This is because the tag of the TLB block does not fit inside the hardware tag entry of the L2 cache.}}
In cases where this condition is not met, an alternative approach is to reduce the number of \konrevd{TLB entries in the TLB Block}
\konrevc{(e.g., \konreve{by} storing 7 PTEs instead of 8 PTEs)} and use the remaining bits for the tag/ASID/VMID. 
Previous works \konrevc{(e.g.,~\cite{graphfire})} propose such solutions to enable efficient sub-block tagging \konreve{in data caches}.


\head{TLB-aware Cache Replacement Policy}
We extend the conventional state-of-the-art SRRIP cache replacement policy~\cite{srrip} to prioritize storing TLB entries \konrevc{of an application} for longer time periods if the application experiences
high address translation overheads (\konrevc{i.e.,} L2 TLB MPKI greater than 5). Listing~\ref{list:srrip} shows the pseudocode of the block insertion function, replacement candidate function, 
and cache hit function for SRRIP in the baseline system and \system (changes compared to baseline SRRIP are marked in blue). 
Upon insertion of a TLB entry inside the L2 cache \nb{(\texttt{insertBlockInL2(block)} Line 1)}, the re-reference interval (analogous to reuse distance) 
is set to 0 (Line 6), marking the TLB entry as a block with a small reuse distance. This way, TLB entries are unlikely to be evicted soon after their insertion. 
\konrevc{Upon selection of a replacement candidate (\texttt{chooseReplacementCandidate()} Line 10), if the selected replacement candidate is a TLB block (Line 23) and translation 
pressure is high (Line 23), SRRIP makes one more attempt to find a replacement candidate that is \emph{not} a TLB block (Line 23).
If no such candidate is found, the TLB block is evicted from the L2 cache and \konrevd{is dropped (i.e., not written anywhere else)}.}
Upon a cache hit to a TLB entry \nb{(\texttt{updateOnL2CacheHit(index)} Line 28)}, the re-reference interval is reduced by three instead of one (Line 32) to provide higher 
priority to the TLB entry compared to other data blocks (Line 34).



\setcounter{listing}{0}

\begin{listing}
\begin{lstlisting}[style=customc]
    function *@\textbf{insertBlockInL2(block)}@*:
        // If inserting a TLB block and TLB pressure is high 
        // set the reference-interval to 0 to provide high priority
        // assuming that reuse in near future will be high
        *@\textcolor{blue}{\textbf{if (block == TLB and TLB\_MPKI > 5)}}@*
            *@\textcolor{blue}{rrip\_counter[block] = 0}@*
        else 
            rrip_counter[block] = RRIP_MAX

    function *@\textbf{chooseReplacementCandidate()}@*:
        // Replace an invalid block if possible
        for i from 0 to m_associativity - 1:
            if (block[i] == invalid ) return i
        // Check the re-reference interval of each block in the set
        for j from 0 to RRIP_MAX:
            // Search for a block with RRIP_MAX
            for i from 1 to #ways:
                chosen_block = chooseBlockWithHighRRIP()
                // If a TLB block is chosen for replacement
                // and translation pressure is high, 
                // make one more attempt and try
                // to evict another block if possible
                *@\textcolor{blue}{\textbf{if (chosen\_block == TLB \&\& TLB\_MPKI > 5) skip}}@* 
            // Increment all RRIP counters 
            for i from 1 to #ways :
                incrementRRIPcounters()

    function *@\textbf{updateOnL2CacheHit(index)}@*:
        // Assume reuse will be high for TLB block
        // and reduce the re-reference interval by 3 instead of 1
        *@\textcolor{blue}{\textbf{if (block[index] == TLB \&\& TLB\_MPKI > 5)}}@*
            *@\textcolor{blue}{\textbf{rrip\_counter[index] -= 3;}}@*
        else
            rrip_counter[index]--;
\end{lstlisting}
\vspace{-2mm}
\caption{TLB-Block-Aware SRRIP~\cite{srrip} L2 Cache Replacement Policy}
\vspace{-2mm}
\label{list:srrip}

\end{listing}

\subsection{\konrevc{Inserting} TLB Blocks \konrevd{into} the L2 Cache}
\label{sec:implementation-alloc-native}
\begin{figure}[h!]
    \centering
    \includegraphics[width=\linewidth]{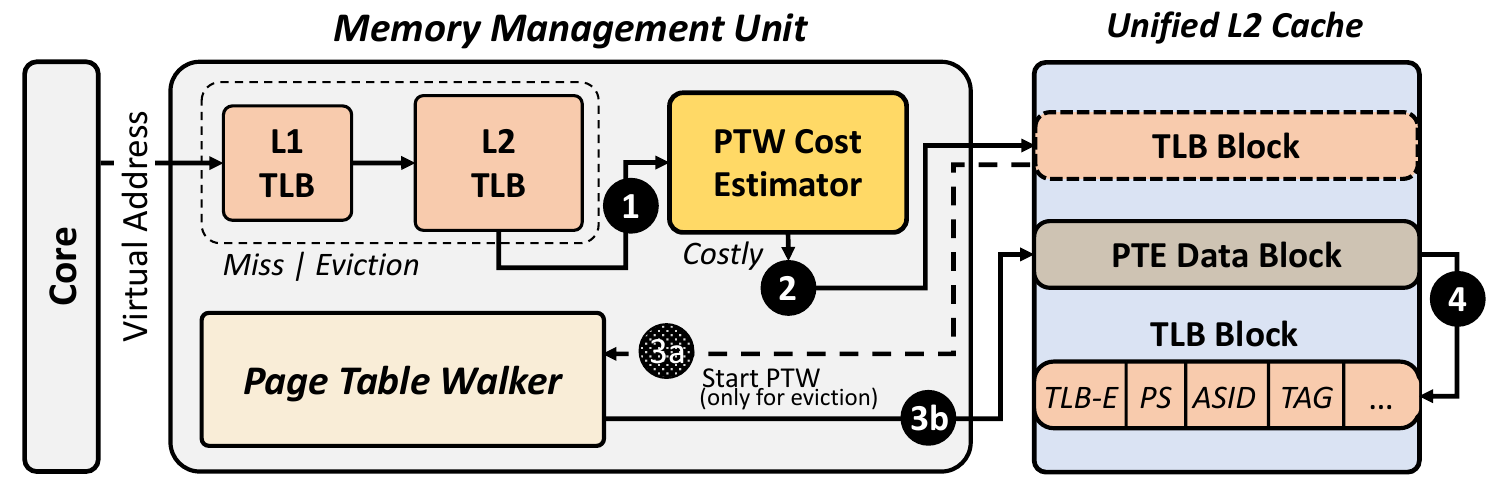}
    \vspace{-6mm}
    \caption{Insertion of a \konrevd{TLB block into} the L2 cache upon (i) an L2 TLB miss and (ii) an L2 TLB eviction.}
    \label{fig:allocation_native}
    \vspace{-2mm}
\end{figure}

\system allocates a \konrevc{block} of 8 TLB entries (64 bytes) that correspond to 8 contiguous virtual pages inside the L2 cache \konrevc{upon} an L2 TLB miss or an L2 TLB eviction,
if the corresponding page \konrevd{is deemed to} be costly-to-translate in the future. To \konrevd{predict} whether a page \konrevc{will be costly-to-translate}, \system employs a Page Table Walk cost predictor (PTW-CP). 
\nb{\autoref{fig:allocation_native} depicts \system's operations on an L2 TLB miss or eviction. }

\head{\konrevc{Inserting a TLB Block into the  L2 Cache upon an L2 TLB Miss}}
When an L2 TLB miss occurs, the MMU consults the PTW-CP to find out if the page \konrevd{is predicted to} be costly-to-translate in the future \nb{(\circled{1} in Fig.~\ref{fig:allocation_native})}. 
If the prediction is positive, the MMU checks if the corresponding TLB block already resides inside the L2 cache \nb{\circled{2}}. 
If it does, no further action is needed. If not, the MMU first waits until the PTW is completed \nb{\circled{3b}}.
When the last level of the PT is fetched, the MMU transforms the cache block that contains the PTEs to a TLB block by updating 
the metadata of the block~\nb{\circled{4}}. The MMU (i) replaces the existing tag with the tag of the virtual page region, (ii) 
sets the TLB bit to mark the cache block as TLB block, and (iii) updates the ASID and the page size information \konrevd{associated with the TLB block}. 
This way, the TLB block containing the \konrevc{consecutive} PTE entries is directly accessible using 
the \konrevc{corresponding} virtual page numbers and the ASID of the application \emph{without} walking the PT. 
Storing \konreve{several (e.g., 8 in our implementation)} TLB entries for \konrevd{consecutive} virtual pages inside the \konreve{same} L2 cache \konreve{TLB block} 
can be highly beneficial for applications whose memory acccesses exhibit high spatial locality and frequently access neighboring pages.

\head{\konrevc{Inserting a TLB Block into the L2 Cache upon an L2 TLB Eviction}}
When an L2 TLB eviction occurs, the MMU consults the PTW-CP to find out if the page \konrevd{is predicted to} be costly-to-translate in the future~\nb{(\circled{1} in Fig.~\ref{fig:allocation_native})}. 
If the outcome of the prediction is positive, the MMU checks if the corresponding TLB block already resides in the L2 cache~\nb{\circled{2}}. 
If it does, no further action is needed. If it does not, the MMU issues in the background a PTW for the corresponding TLB entry~\nb{\circled{3a}}. 
When the last level of the page table is fetched~\nb{\circled{3b}}, the MMU follows the same procedure as the L2 TLB miss-based insertion \konrevd{(i.e., transforms the cache block that contains the PTEs to a TLB block)}~\circled{4}.
This way, if the evicted TLB entr\konreva{y} (or any other TLB entry in the block) \konreva{is} accessed again in the future, 
\system can directly access the corresponding PTE without walking the PT.

\begin{figure}[h!]
    \vspace{-2mm}
    \centering
    \includegraphics[width=\linewidth]{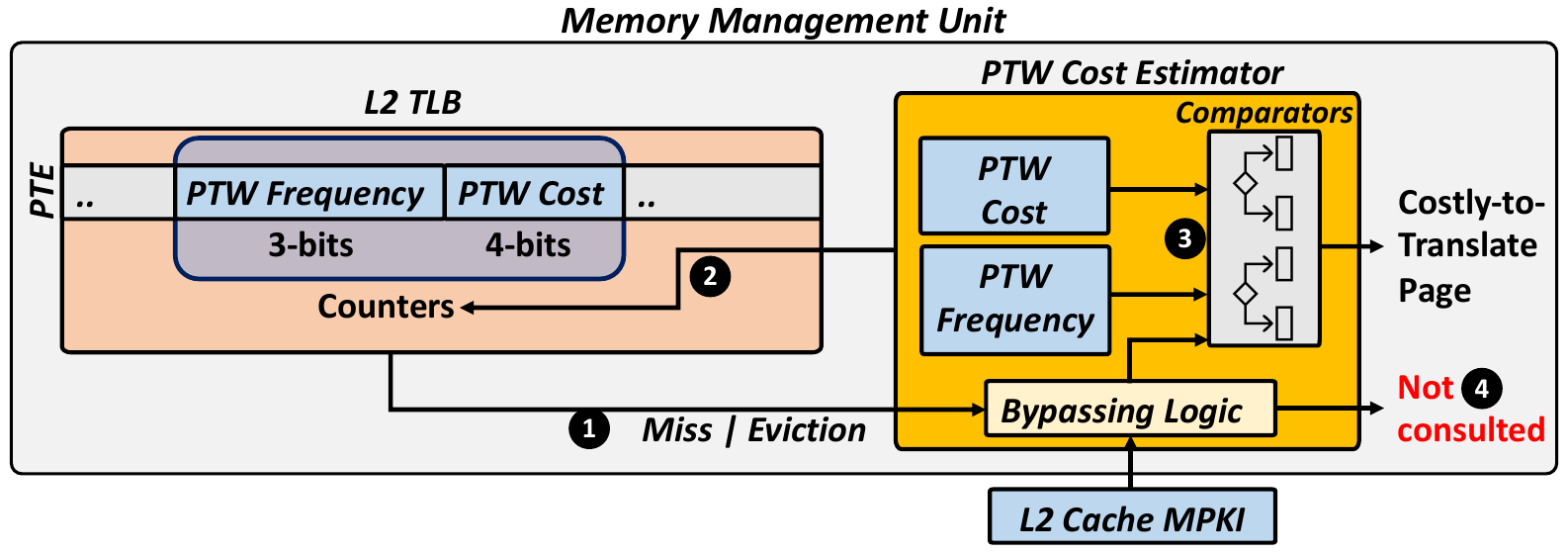}
    \vspace{-7mm}
    \caption{Page Table Walk Cost Predictor.}
    \label{fig:ptwcp}
    \vspace{-2mm}
\end{figure}

\head{Page Table Walk Cost Predictor: Functionality}
The PTW cost predictor (PTW-CP) is a \konrevc{small} \konrevc{comparator-based circuit that} estimates whether the page is among the top 30\% most costly-to-translate pages. \konrevc{Using it,} 
\system \konrevd{predicts} if \konrevc{a} page will cause costly PTWs in the future and decides whether the MMU should store the corresponding TLB block inside the L2 cache.
To make this decision, PTW-CP uses two metrics~\konrevc{associated with a page}: (i) PTW frequency and (ii) PTW cost, \konrevd{both of} which are embedded inside the PTE of the corresponding page. 
Figure~\ref{fig:ptwcp} shows the structure and the functionality of PTW-CP.
PTW frequency is stored as a 3-bit counter in the unused bits of \konrevc{the} PTE and is incremented after every PTW that fetches the corresponding PTE. 
PTW cost is also stored \konrevc{as a 4-bit counter} in the unused bits of the PTE and is incremented every time the PTW \konrevc{leads} to at least one DRAM access.
\konreve{Both counters are updated \konrevd{by the MMU} \konrevc{after} every PTW that fetches the corresponding PTE.}
\konreve{If any of the two counters overflows, its value remains at the maximum value throughout the rest of the program's execution.}

On an L2 TLB miss or eviction~\circled{1}, the PTW-CP waits until the corresponding PTE is fetched inside the L2 TLB. 
PTW-CP fetches the two counters~\circled{2} from the TLB entry that contains the PTE, passes them through a tree of comparators, and calculates the result~\circled{3}.
If the L2 cache experiences high MPKI \konrevd{(i.e., data exhibits low locality, meaning that caching data is not that beneficial)}, 
the PTW-CP is bypassed and the TLB entry is inserted inside the L2 cache without consulting the PTW-CP~\circled{4}.

\head{Page Table Walk Cost Predictor: Feature Selection}
Our development of PTW-CP's architecture involves a systematic and empirical approach to (i) identify the most critical features for making high-accuracy predictions and
(ii) create an effective predictor while minimizing hardware overhead and inference latency.
Initially, we collect a set of 10 per-page features related to address translation, as shown in Table~\ref{tab:features}.
\konrevd{From} these 10 features, we \konrevd{methodically} identify a \konrevd{small} subset that would maximize accuracy while minimizing \konrevd{prediction} time and storage overhead.


\begin{table}[!ht]
    \centering
    \scriptsize
    \caption{\konkanelloreve{Per-Page Feature Set}}
    \vspace{-4mm}
    \begin{tabular}{|m{12.5em}|m{1em}m{20.5em}|}
    \hline
        \textbf{Feature (per PTE)} & \textbf{Bits} & \textbf{Description} \\ \hline
        Page Size & 1 & The size of the page (4KB or 2MB) \\ 
        \textbf{Page Table Walk Frequency} & \textbf{3} & \textbf{\# of PTWs for the page} \\ 
        \textbf{Page Table Walk Cost} & \textbf{4} & \textbf{\# of DRAM accesses during all PTWs}\\ 
        PWC Hits & 5 & \# of times the PTW led to PWC hit \\ 
        L1 TLB Misses & 5 & \# of times the page experienced L1 TLB miss \\ 
        L2 TLB Misses & 5 & \# of times the page experienced L2 TLB miss \\ 
        L2 Cache Hits & 5 & \# of times the page experienced L2 cache hits \\ 
        L1 TLB Evictions & 5 & \# of times the TLB entry got evicted from L1 TLB \\ 
        L2 TLB Evictions & 6 & \# of times the TLB entry got evicted from L2 TLB \\ 
        Accesses & 6 & \# of accesses to the page \\ \hline
    \end{tabular}
    \label{tab:features}
    \vspace{-1mm}

\end{table}

\konrevd{Table \ref{tab:ptw_perf} shows the architectural characteristics and the performance of three different multi-layer perceptron-based neural networks (NN)~\cite{haykin1994neural} and of \konreve{our} final comparator-based model.}
First, we evaluate three different NN architectures with different feature sets to gain insights about the most critical features (\konrevd{for accuracy}).
The first NN (\texttt{NN-10}) uses all 10 features, the second NN (\texttt{NN-5}) uses a set of 5 features (PTW cost, PTW frequency, PWC hits, L2 TLB evictions, and accesses to the page), 
and the third (\texttt{NN-2}) uses only 2 features, the PTW frequency and the PTW cost.
We use four metrics to evaluate the performance of each model: accuracy, precision, recall, and F1-score.
Accuracy \konrevd{is the fraction} of correct predictions, precision \konrevd{is the fraction} of correct positive \konrevd{(i.e., costly-to-translate)} predictions, 
and recall \konrevd{is the fraction} of correct negative predictions. F1-score is the harmonic mean of precision and recall.
In the context of PTW-CP, making negative predictions when the page is \konrevd{actually} costly-to-translate leads to performance degradation, 
while making positive predictions when the page is \konrevd{actually} \emph{not} costly-to-translate leads to L2 cache pollution.
From Table~\ref{tab:ptw_perf}, we observe that NN-10 achieves the highest performance, with an F1-score of 90.42\%. By reducing the number of features to 5, 
NN-5 still achieves high performance reaching 89.89\% F1-score while NN-2 leads to an F1-score of 80.66\%. 
At the same time, NN-2 is 7.75x smaller than NN-10 and 90.5x smaller than NN-5 which makes it an attractive solution for PTW-CP
as it achieves \konrevd{reasonable} accuracy with \konrevd{small} hardware overhead.

\begin{table}[!ht]
        \centering
    \scriptsize
    \caption{\konkanelloreve{Comparison of Different Types of PTW-CP}}
    \vspace{-4mm}
    \begin{tabular}{|m{12em}|m{4em}m{4em}m{4em}m{7em}|}
    \hline
        \textbf{\textit{Model Parameters}} & \texttt{NN-10} & \texttt{NN-5} & \texttt{NN-2} & \textbf{Comparator} \\ \hline
        \# of Features & 10 & 5 & 2 &\textbf{2} \\ 
        Number of Layers & 4 & 4 & 6 & \textbf{N/A} \\ 
        Size of Hidden Layers & 16 & 64 & 4 & \textbf{N/A} \\ 
        Size (B) & 6024 & 70152 & 776 & \textbf{24} \\ 
        Recall & 93.34\% & 92.44\% & 89.62\% & \textbf{89.61\%} \\ 
        Accuracy & 92.13\% & 91.72\% & 82.90\% & \textbf{82.90\%} \\ 
        Precision & 87.68\% & 87.47\% & 73.33\% & \textbf{73.34\%} \\ 
        F1-score & 90.42\% & 89.89\% & 80.66\% & \textbf{80.66\%} \\ \hline
    \end{tabular}
    \label{tab:ptw_perf}
\end{table}

\begin{figure}[h!]
    \vspace{-2mm}
    \centering
    \includegraphics[width=\linewidth]{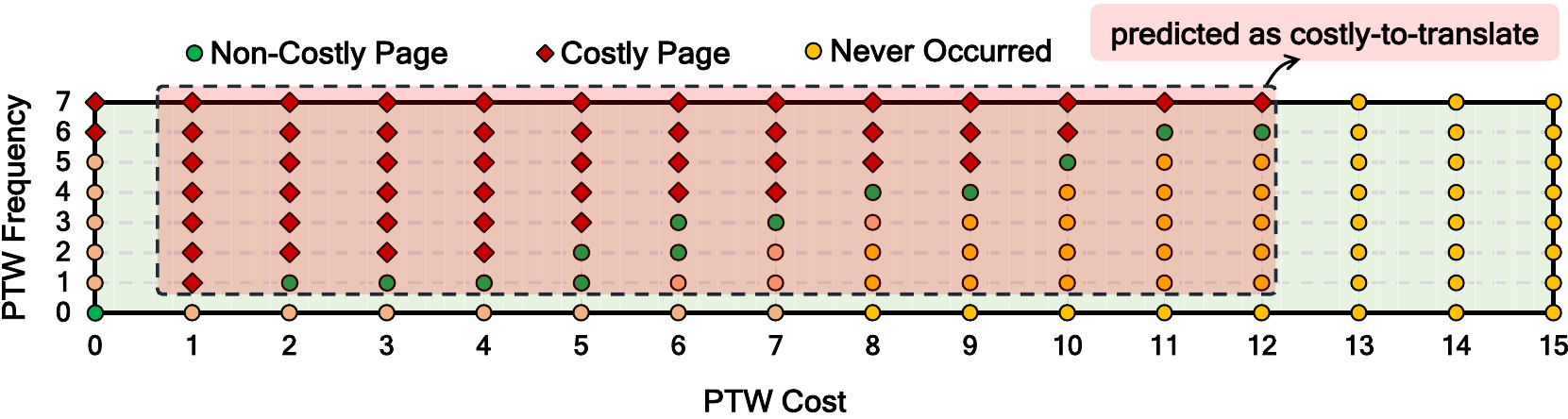}
    \vspace{-6mm}
    \caption{\konkanelloreve{Prediction pattern of NN-2. The bounding box separates the PTW cost-frequency pairs that lead to 
    positive predictions (inside the box) from the ones that lead to negative predictions (outside the box).}}
    \label{fig:scatter}
    \vspace{-4mm}
\end{figure}

\konrevc{To gain a better understanding of the prediction pattern of NN-2, Fig.~\ref{fig:scatter} shows the predictions of the network for all possible PTW frequency and PTW cost value pairs. 
We observe that the network exhibits a clear prediction pattern that separates costly-to-translate pages from non-costly-to-translate pages: 
PTW frequency-cost value pairs that fall inside the boundaries of the bounding box (rectangle spanning from the bottom-left corner (1,1) to the top-right corner (12,7) as drawn on Fig.~\ref{fig:scatter}) 
are classified as costly-to-translate by NN-2, while PTW frequency-cost value pairs that fall outside the bounding box are classified as non-costly-to-translate. \konreve{Many of the PTW frequency-cost value pairs
never occur during the execution of the applications we evaluate and are not classified by NN-2.}
Table~\ref{tab:ptw_perf} demonstrates that a simple comparator approach that mimics the functionality the bounding box shown in Fig.~\ref{fig:scatter}, 
achieves an F1-score of 80.66\% without any performance loss compared to NN-2. The comparator-based model requires only 24 bytes of storage,
251x less than NN-10, 2923x less than NN-5 and 32x less than NN-2. The comparator-based model requires only 
(i) four comparators to compare the two counters with the edges of the bounding box, i.e., (1,1) and (12,7)} and (ii) can make a prediction in a single cycle.
The comparator-based model is the PTW-CP architecture that we use in \system.



\subsection{Address Translation Flow with \system}

\begin{figure}[h!]
    \centering
    \includegraphics[width=\linewidth]{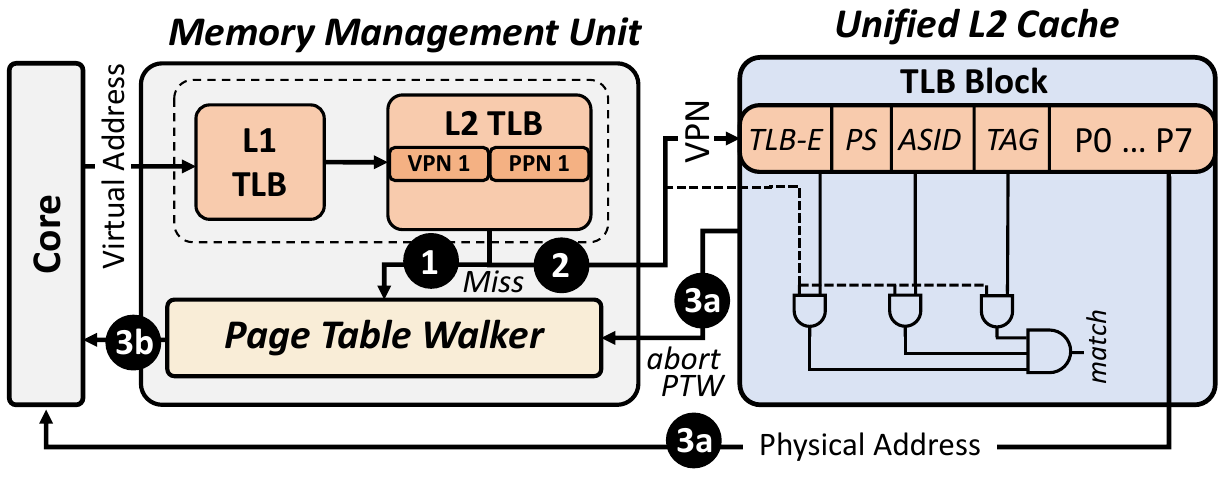}
    \vspace{-5mm}
    \caption{Address translation flow in a system with Victima.}
    \label{fig:translation_native}
    \vspace{-2mm}
\end{figure}

\autoref{fig:translation_native} demonstrates the address translation flow in a system that employs \system. 
When \konrevd{an} L2 TLB miss occurs, the MMU in parallel (i) initiates the PTW~\circled{1} and (ii) \konrevd{looks up} \nb{the corresponding TLB block in} the L2 cache~\nb{\circled{2}}.
In contrast \nb{to} regular L2 data block lookups, which are performed using the physical address, a TLB block lookup is performed using the virtual page number (VPN) and the address-space identifier (ASID) of the translation request.
The size of the VPN is not known a priori, so \system probes the L2 cache twice in parallel, once assuming a 4KB VPN and once assuming a 2MB VPN.
If the tag (either the tag of \konrevd{the} 4KB VPN or the tag of the 2MB VPN) and the ASID matches with a block that has the TLB-entry bit set, 
the translation request is served by the L2 cache~\nb{\circled{3a}}, the PTW is aborted, and the TLB entry is inserted into the L2 TLB.
\konrevd{If} the TLB entry is not found in the L2 cache, the PT walker \konrevd{runs to completion and} resolves the translation~\nb{\circled{3b}}.

\subsection{\system in Virtualized Environments}

\begin{figure}[h!]
    \centering
    \includegraphics[width=\linewidth]{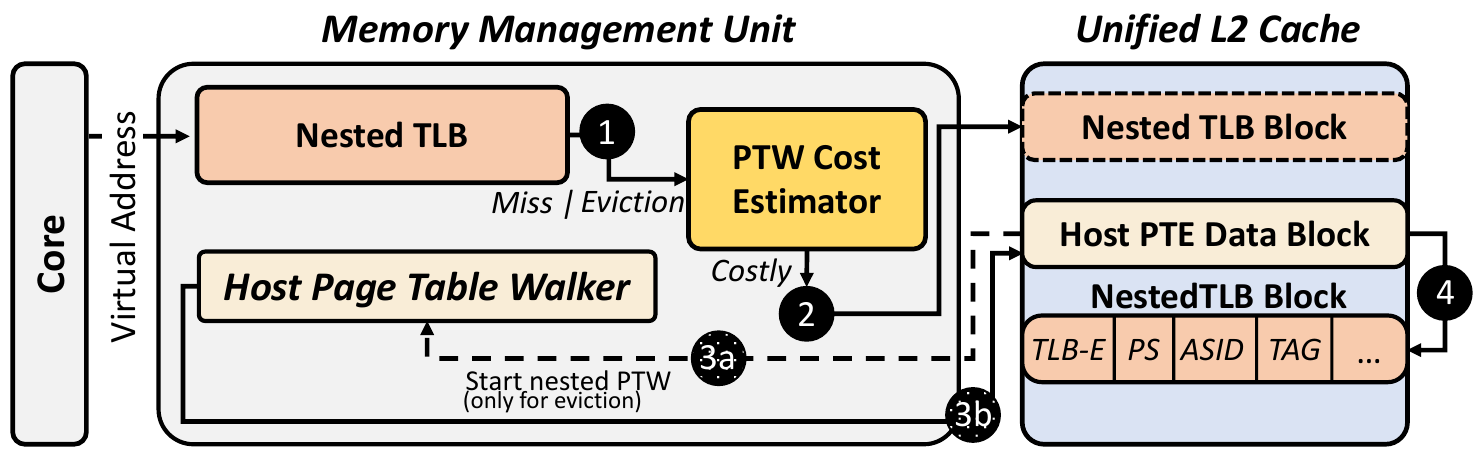}
    \vspace{-4mm}
    \caption{Insertion of a nested TLB block into the L2 cache upon (i)
    a nested TLB miss and (ii) a nested TLB eviction}
    \label{fig:victima_virt}
    \vspace{-2mm}
\end{figure}

\konrevd{We} demonstrate how \system improves address translation in virtualized environments. 
The key idea is to insert both (i) TLB entries and (ii) \emph{nested} TLB entries into the L2 cache to 
increase the translation reach of the processor's TLB hierarchy for both guest-virtual-to-guest-physical and 
guest-physical-to-host-physical \konrevd{address} translations and avoid both (i) guest-PTWs and (ii) host-PTWs. 
A nested TLB block is a block of 8 nested TLB entries that correspond to 8 contiguous host-virtual pages. 
\konreve{To distinguish between conventional TLB blocks and nested TLB blocks, \system extends the cache block metadata with an additional bit to mark a block as a nested TLB block.}
\autoref{fig:victima_virt} shows how Nested TLB blocks are inserted into the L2 cache in a system that employs \system and nested paging~\cite{amdnested} in virtualized execution. 
(conventional TLB blocks are allocated as described in \S\ref{sec:implementation-alloc-native}).

\head{\konrevd{Inserting a Nested TLB Block into the L2 Cache upon a Nested TLB Miss}}
\konrevd{When a \konrevd{Nested} TLB miss occurs, the MMU consults the PTW-CP to find out if the host-virtual page will be costly-to-translate in the future \nb{\circled{1}}. 
If the prediction is positive, the MMU checks if the corresponding nested TLB block already resides inside the L2 cache \nb{\circled{2}}. 
If it does, no further action is needed. If not, the MMU first waits until the host-PTW is completed.
When the last level of the host-PT is fetched  \nb{\circled{3b}}, the MMU transforms the cache block that contains the host-PTEs to a nested TLB block by updating 
the metadata of the block~\nb{\circled{4}}. The MMU (i) replaces the existing tag with the tag of the host-virtual page region, (ii) 
sets the nested TLB bit to mark the cache block as a nested TLB block, and (iii) updates the ASID (or VMID) and the page size information.}

\head{\konrevd{Inserting a Nested TLB Block into the L2 Cache upon a Nested TLB Eviction}}
\konrevd{When a \konrevd{Nested} TLB eviction occurs, the MMU consults the PTW-CP to find out if the host-virtual page will be costly-to-translate in the future~\nb{\circled{1}}}. 
If the outcome of the prediction is positive, the MMU checks if the corresponding nested TLB block already resides in the L2 cache~\nb{\circled{2}}. 
If it does, no further action is needed. If it does not, the MMU issues in the background a host-PTW for the corresponding TLB entry~\nb{\circled{3a}}. 
When the last level of the host-PT is fetched~{\circled{3b}}, the MMU transforms the cache block that contains the host-PTEs to a nested TLB block.
~\nb{\circled{4}}.

\head{Address Translation Flow}
\nb{\autoref{fig:victima_virt} shows the address translation flow of a system that employs \system and nested paging~\cite{amdnested} in virtualized execution. }
If a nested TLB miss occurs \nb{\circled{1}}, the MMU probes the L2 cache to search for the nested TLB entry \nb{\circled{2}}. 
If the nested TLB entry is found inside the L2 cache, the host-PTW gets skipped~\nb{\circled{3a}}. 
If it is not found, the host-PTW is performs the guest-physical-to-host-physical address translation \nb{\circled{3b}}.

\begin{figure}[h!]
    \centering
    \includegraphics[width=\linewidth]{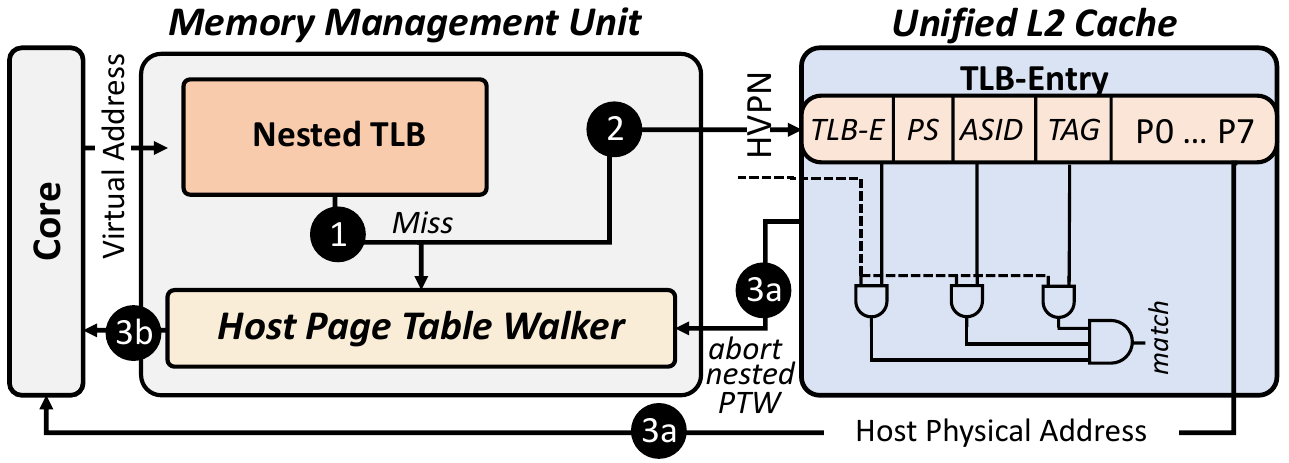}
    \vspace{-6mm}
    \caption{Address translation flow in a system with Victima in a virtualized execution environment.}
    \vspace{-4mm}
\end{figure}

\section{TLB Maintainance Operations}
\label{sec:coherence}
\konkanellorevd{Modern ISAs provide specific instructions used by the OS to invalidate TLB entries and maintain correctness \konrevd{in the presence of} 
(i) context switches and (ii) modifications of virtual-to-physical address mappings (called TLB shootdowns) \konrevd{that} occur \konrevd{due to} physical page migration, memory de-allocation  etc.
Different ISAs provide different instructions for TLB invalidations.
For example, the ARM v8 architecture~\cite[D5.10.2]{arm-manual-tlbmaintenance} defines \konrevd{multiple} special instructions to invalidate TLB entries 
\konrevd{with each instruction handling a distinct case (e.g., invalidating a single TLB entry vs invalidating all TLB entries with a specific ASID).}
x86-64 provides a single instruction, \texttt{INVLPG}, which corresponds to invalidating one single TLB entry~\cite{intelx86manual-vol2}.
In Victima, whenever a TLB invalidation is required, the corresponding TLB entries in the L2 cache need to be invalidated.  
In this section, following the example of the ARM specification, which is a superset of other 
specifications \konreve{we know of}, we discuss in detail how Victima supports TLB invalidations \konrevd{due to} context-switches and TLB shootdowns.}

\subsection{Context Switches}
TLB flushing occurs when the OS \konrevd{switches the hardware context} and schedules another process \konrevd{(or thread)} to the core. 
In this case, the OS makes a decision on whether \konrevd{or not} the TLB entries across the TLB hierarchy \konrevd{should} be invalidated, 
which depends on the ASIDs of the current and to-be-executed processes (in practice Linux uses only 12 different ASIDs per core even though the processor can support up to 4096 ASIDs). 
In Victima, if the OS \konrevd{flushes the \konreve{entire} TLB hierarchy}, \konrevd{all the TLB blocks in L2 cache need to be invalidated as well.}
If the OS performs a partial flush based on the ASID, all the TLB blocks in L2 cache with the corresponding ASID need to be evicted. 
In the corner case that Victima uses fewer bits for the ASID, i.e., when L2 cache tag is not large enough to store \konrevd{enough} ASID bits 
\konrevd{to cover} the ASID of the process, all the TLB blocks inside the L2 cache get invalidated during a context switch 
(the L1 and L2 TLB entries can still be invalidated using the ASID). Based on our evaluation setup, for a 2MB L2 cache which is occupied by 50\% by TLB blocks, 
the total time to complete the invalidation procedure is \konrevd{on} the order of 100 ns.  The invalidation procedure happens in parallel with the L2 TLB 
invalidation and is negligible compared to context switch completion times (order of $\mu$s\VMcontextswitch).

\head{(i) \konrevd{Invalidating} all TLB entries} To invalidate all the TLB blocks inside the L2 cache, the L2 TLB first sends an invalidation command to the L2 cache controller. 
The cache controller \konrevd{probes in parallel} all cache banks to invalidate all the TLB blocks of every L2 \konrevd{cache} set. 
For each way, if the TLB entry bit is set, the TLB block is invalidated.

\head{(ii) \konrevd{Invalidating} all TLB entries with a specific ASID} To invalidate all TLB blocks with a specific ASID, 
the L2 TLB first sends an invalidation command to the L2 cache controller with the corresponding ASID. 
For every cache block, if the TLB entry bit is set and the ASID matches the ASID of the invalidation request, the TLB block is invalidated.
If the size of the ASID of the invalidation command is larger (e.g., 4 bits) than the supported ASID (e.g., 3 bits), then all the TLB blocks inside L2 cache are flushed.
\konreve{However, we believe this is an uncommon case, because, e.g., Linux uses only 12 ASIDs/core~\cite{linuxasid}.}

\subsection{TLB Shootdowns}

\konkanellorevd{A TLB shootdown occurs when the CPU needs to invalidate stale TLB entries on local and remote cores. 
\konrevd{It is caused by various} memory \konrevd{management} operations that modify page table entries, such as de-allocating pages (unmap()), 
migrating pages, page permission changes, deduplication,  and memory compaction. As shown in previous works~\cite{latr}, TLB shootdowns take order of $\mu$s time 
to complete \konrevd{due to} expensive inter-processor interrupts (IPIs). In Victima, if the system performs a TLB shootdown, the corresponding TLB blocks need to be 
invalidated in the L2 cache. We explain how for two different TLB shootdown-based invalidations:}

\konkanellorevd{\head{(i) \konrevd{Invalidating a single TLB entry} given VA and ASID} Invalidating a specific TLB entry by VA and ASID 
only requires sending an invalidation command with the VA and the ASID to the L2 cache controller. 
Since each TLB block contains eight contiguous TLB entries, 
invalidating one TLB entry of the TLB block leads to invalidating all eight corresponding TLB entries.}

\konkanellorevd{\head{(ii) \konrevd{Invalidating all TLB entries given a range of VAs}} Invalidating a range of VAs requires
sending multiple invalidation commands with different VAs to the L2 cache controller. The L2 cache controller 
accordingly invalidates all the corresponding TLB blocks.}


\section{Area \& Power Overhead}
\label{sec:overheads}
\konrevd{
\system requires \konrevd{three} \konrevd{additions} to an existing high-performance core design: 
(i) \konrevd{two new \emph{TLB Entry} bits in every L2 cache block (one of TLB entries and one for nested TLB entries)} (\S\ref{sec:l2-changes}), 
(ii) the PTW cost estimator (\S\ref{sec:implementation-alloc-native}) and (iii) the necessary logic to perform tag matching \konreve{and} invalidation of TLB blocks using the \emph{TLB Entry} bit, the VPN, and the ASID (\S\ref{sec:coherence}).
Extending each L2 cache block with two \emph{TLB Entry} bits results in a $0.4\%$ storage overhead for caches with $64$B blocks 
(e.g., in total $8$KB for a $2$MB L2 cache). 
PTW-CP requires only (i) 4 comparators to compare the PTE counters with the corresponding thresholds and (ii) 4 registers to store the thresholds.
To support tag matching/invalidation operations for TLB blocks, we extend the tag comparators of the L2 cache with a bitmask to distinguish between tag matching/invalidation for TLB blocks and tag matching/invalidation for conventional data blocks.
Based on our evaluation with McPAT~\cite{mcpat}, all additional logic requires $0.04\%$ area overhead and $0.08\%$ power overhead on top of the high-end Intel Raptor Lake processor~\cite{raptor_lake}.
}

\section{Evaluation Methodology}

\label{sec:methodology}

We evaluate \system using an extended version of the Sniper Multicore Simulator \cite{sniper}.
\konrevd{This simulator and its documentation are freely available at \textcolor{blue}{\url{https://github.com/CMU-SAFARI/Victima}}.}
We extend Sniper to accurately model: (i) TLBs that support multiple page sizes, (ii) the conventional radix page table walk, (iii) page walk caches, (iv) nested TLBs
and nested paging~\cite{amdnested} and, (vi) the functionality and timing of all the evaluated systems.
Table \ref{tab:config} shows the simulation configuration of (i) the baseline system and (ii) all evaluated systems.

\definecolor{SoftPeach}{rgb}{0.937,0.901,0.901}
\begin{table}[h!]
\scriptsize
\centering
\caption{Simulation Configuration and Simulated Systems}
\vspace{-2mm}
\label{tab:simconfig}
\begin{tblr}{
  width = \linewidth,
  colspec = {Q[237]Q[704]},
  row{1} = {SoftPeach,c},
  row{13} = {SoftPeach,c},
  cell{1}{1} = {c=2}{0.941\linewidth},
  cell{3}{1} = {r=4}{},
  cell{7}{1} = {r=2}{},
  cell{9}{1} = {r=2}{},
  cell{13}{1} = {c=2}{0.941\linewidth},
  cell{14}{1} = {r=2}{},
  cell{16}{1} = {r=2}{},
  cell{20}{1} = {r=2}{},
  cell{22}{1} = {r=2}{},
  cell{24}{1} = {r=2}{},
  vlines,
  hline{1-3,12-14,19-20,22,24} = {-}{},
  hline{4-11,15-18,21,23,25-26} = {-}{},
}
\textbf{Baseline System} & \\
\textbf{Core} & 4-way OoO x86-64 2.6GHz \\
\textbf{MMU} & L1 I-TLB: 128-entry, 8-way assoc, 1-cycle latency\\
 & {L1 D-TLB (4KB): 64-entry, 4-way assoc, 1-cycle latency \\ L1 D-TLB (2MB): 32-entry, 4-way assoc, 1-cycle latency}\\
 & L2 TLB: 1536-entry, 12-way assoc, 12-cycle latency\\
 & 3 Split Page Walk Caches: 32-entry, 4-way assoc,  2-cycle latency\\
\textbf{L1 Cache} & {L1 I-Cache: 32~KB, 8-way assoc, 4-cycle access latency \\ L1 D-Cache: 32~KB, 8-way assoc, 4-cycle access latency}\\
 & LRU replacement policy;~IP-stride prefetcher~\cite{stride}\\
\textbf{L2 Cache} & 2~MB, 16-way assoc, 16-cycle latency\\
 & SRRIP replacement policy~\cite{srrip}; Stream prefetcher~\cite{streamer}\\
\textbf{L3 Cache} & 2~MB/core, 16-way assoc, 35-cycle latency\\
\textbf{\textbf{Transparent \newline Huge Pages~\cite{arcangeli2010,corbet2011}}} & 
Debian 9 4.14.2. 10-node cluster \newline Memory per node: 256GB-1TB\\
\textbf{\textbf{Evaluated Systems}} & \\
\textbf{POM-TLB~\cite{pomtlbISCA2017}} & 64K-entry L3 software-managed TLB, 16-way assoc\\
 & TLB-aware SRRIP replacement policy (\S\ref{sec:l2-changes})\\
\textbf{Opt. L3 TLB-64K} & 1.5K-entry L2 TLB, 12-cycle latency\\
 & 64K-entry L3 TLB, optimistic 15-cycle latency\\
\textbf{Opt. L2 TLB-64K} & 64K-entry L2 TLB, 16-way assoc, optimistic 12-cycle latency\\
\textbf{Opt. L2 TLB-128K} & 128K-entry L2 TLB, 16-way assoc,  optimistic 12-cycle latency\\
\textbf{Nested Paging~\cite{amdnested}} & 2D PTW; Guest PT: Four-level Radix, Host PT: Four-level Radix ~\\
 & 64-entry Nested TLB, 1-cycle latency\\
 \textbf{Ideal Shadow Paging (I-SP)~\cite{vm25}} & 1D Shadow PTW instead of 2D PTW \\
 & Updates to shadow page table cause no performance overheads\\
\textbf{Victima} & MMU consults PTW-CP only if L2 cache MPKI $<$ 5 (\S\ref{sec:implementation-alloc-native})\\
 & TLB-aware SRRIP replacement policy (\S\ref{sec:l2-changes})
\end{tblr}
\label{tab:config}
\end{table}

\head{Workloads} Table~\ref{tab:workloads} shows all the benchmarks we use to evaluate Victima and the systems we compare Victima to.
We select applications with high L2 TLB MPKI ($>5$), which are also used in previous works~\cite{elastic-cuckoo-asplos20,compendiaISMM2021,midgard,flataAsplos2022}. 
We evaluate our design using seven workloads from the GraphBig \cite{Lifeng2015} suite, XSBench \cite{Tramm2014},
the Random access workload from the GUPS suite~\cite{Plimpton2006}, Sparse Length Sum from DLRM~\cite{dlmr} and kmer-count from GenomicsBench~\cite{genomicsbench}. 
We extract the page size information for each workload from a real system that uses Transparent Huge Pages~\cite{corbet2011,arcangeli2010} with both 4KB and 2MB pages. 
\konrevd{Each benchmark is executed for 500M instructions.}

\begin{table}[ht!]
  \centering
  \scriptsize
  \caption{Evaluated Workloads}
  \vspace{-2mm}
    \begin{tabular}{m{8em}m{20.5em}r}
    \toprule
    \textbf{Suite} & \textbf{Workload} & \textbf{Dataset size} \\
    \midrule
    GraphBIG~\cite{Lifeng2015} & Betweeness Centrality (BC), Bread-first search (BFS), Connected components (CC), Graph coloring (GC), PageRank (PR), Triangle counting (TC), Shortest-path (SP)   & 8 GB \\
    \midrule
    XSBench~\cite{Tramm2014} & Particle Simulation (XS)     & 9 GB \\
    \midrule
    GUPS~\cite{Plimpton2006}  & Random-access (RND) & 10 GB \\
    \midrule
    DLRM~\cite{dlmr}  & Sparse-length sum (DLRM) & 10.3 GB \\
    \midrule
    GenomicsBench~\cite{genomicsbench} & k-mer counting (GEN) & 33 GB \\
    \bottomrule
    \end{tabular}%
  \label{tab:workloads}%

\end{table}%

\head{Evaluated Systems in Native Execution}
\konkanelloreva{We evaluate six different systems in native execution environments}:
(i) \textbf{\textit{Radix}}: Baseline x86-64 system that uses the conventional (1) two-level TLB hierarchy and (2) four-level radix-based page table,
(ii) \textbf{\textit{POM-TLB}}: a system equipped with a large 64K-entry software-managed L3 TLB~\cite{pomtlbISCA2017} and the TLB-aware SRRIP policy (\S\ref{sec:l2-changes}) at the L2 cache, 
(iii) \textbf{\textit{Opt. L3TLB-64K}}: a system equipped with a 64K-entry L3 TLB with an optimistic 15-cycle access latency, 
(iv) \textbf{\textit{Opt. L2TLB-64K}}: a system equipped with a 64K-entry L2 TLB with an optimistic 12-cycle access latency, 
(v) \textbf{\textit{Opt. L2TLB-128K}}: a system equipped with a 128K-entry L2 TLB with an optimistic 12-cycle access latency, and 
(vi) \textbf{\textit{\system}}: a system that employs \system and the TLB-aware SRRIP policy (\S\ref{sec:l2-changes}) at the L2 cache.

\head{Evaluated Systems in Virtualized Execution}
We evaluate four different systems in virtualized execution environments: 
(i) \textbf{\textit{Nested Paging (NP)}}: Baseline x86-64 system that uses (1) a two-level TLB hierarchy and (2) a 64-entry Nested TLB and employs Nested Paging~\cite{amdnested}, 
(ii) \textbf{\textit{POM-TLB}}: a system equipped with a large 64K-entry software-managed L3 TLB~\cite{pomtlbISCA2017} and the TLB-aware SRRIP policy (\S\ref{sec:l2-changes}) at the L2 cache, 
(iii) \textbf{\textit{I-SP}}: a system that employs an ideal version of shadow paging~\cite{amdnested,vm25} where (1) only a four-level radix shadow page table walk is needed to discover the virtual-to-physical translation and 
(2) the updates to the shadow page table are performed without incurring performance overhead, and
(iv) \textbf{\textit{\system}}: a system that employs \system and caches both TLB and nested TLB entries in the L2 cache which is equipped with the TLB-aware SRRIP policy at the L2 cache.

\section{Evaluation Results}
\label{sec:results}
\subsection{Native Execution Environments}
\label{sec:native_results}
 
\begin{figure}[h]
    \vspace{-2mm}
    \centering
    \includegraphics[width=\linewidth]{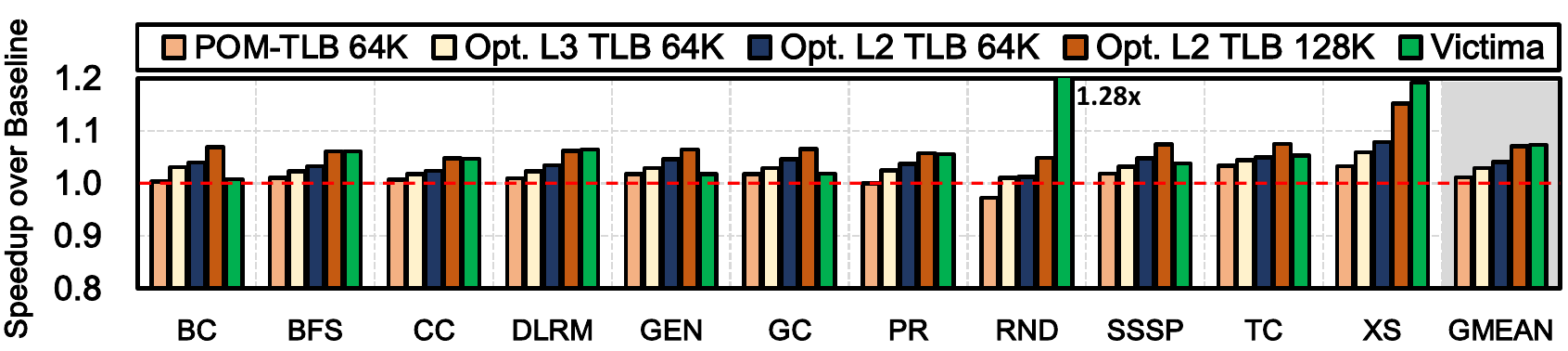}
    \vspace{-6mm}
    \caption{\konkanelloreva{Speedup provided by POM-TLB, Opt. L3TLB-64K, Opt. L2TLB-64K, Opt. L2TLB-128K  and \system over Radix.}}
    \label{fig:speedup_victima}
    \vspace{-1mm}
\end{figure}

\konkanelloreva{\autoref{fig:speedup_victima} shows the execution time speedup \konreve{provided} by POM-TLB, Opt. L3TLB-64K, Opt. L2TLB-64K, Opt. L2TLB-128K and \system compared to Radix.
We make two key observations:
First, \system on average respectively outperforms Radix, POM-TLB, Opt. L3TLB-64K, Opt. L2TLB-64K, by \speedupoverbaseline\%, \speedupoverpomtlb\%, 4.4\%, 3.3\%. 
In RND, which follows highly irregular access patterns, \konreve{\system improves performance by 28\% over Radix.}
Second, Victima achieves similar performance gains as Opt.L2-TLB 128K without the latency/area/power overheads associated with an 128K-entry TLB. 
To better understand the performance benefits achieved by \system, we examine the impact of \system \konreve{on} (i) the number of PTWs and (ii) the L2 TLB miss latency.}

\begin{figure}[t]
    \vspace{-1mm}
    \centering
    \includegraphics[width=\linewidth]{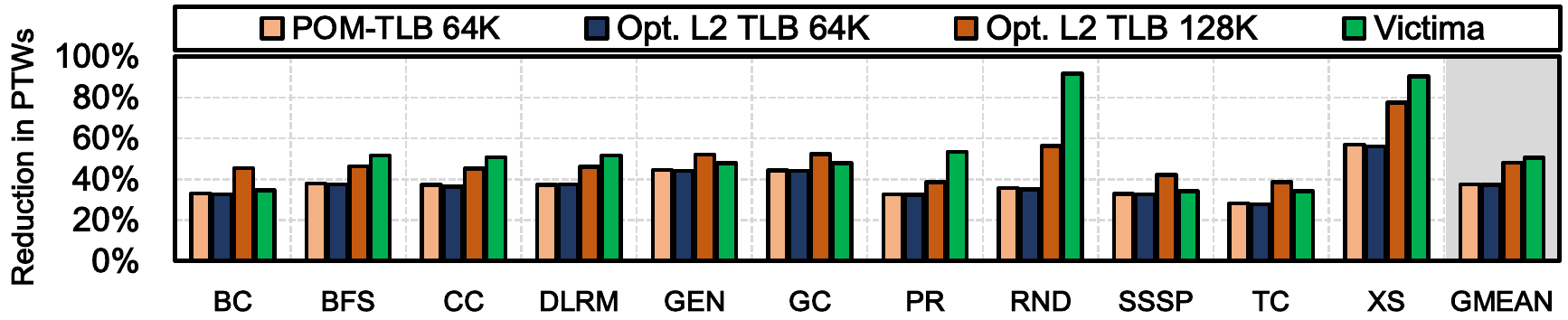}
    \vspace{-6mm}
    \caption{\konkanelloreva{Reduction in PTWs provided by POM-TLB, L2 TLB-64K, L2 TLB-128K, and \system over Radix.}}
    \label{fig:victima_ptws}
    \vspace{-1mm}
\end{figure}

Figures~\ref{fig:victima_ptws} shows the reduction in PTWs achieved by POM-TLB, L2 TLB-64K, L2 TLB-128K and \system over Radix, in a native execution environment, across 11 workloads.
We observe that \system reduces the number of PTWs by $50\%$, POM-TLB by $37\%$, L2 TLB-64K by $37\%$ and L2 TLB-128K by $48\%$ on average across all workloads. 
L2 TLB-128K and Victima lead to similar reductions \konreve{in} PTWs, which explains the similar performance gains of the two mechanisms.

\begin{figure}[h]
    \vspace{-2mm}
    \centering
    \includegraphics[width=\linewidth]{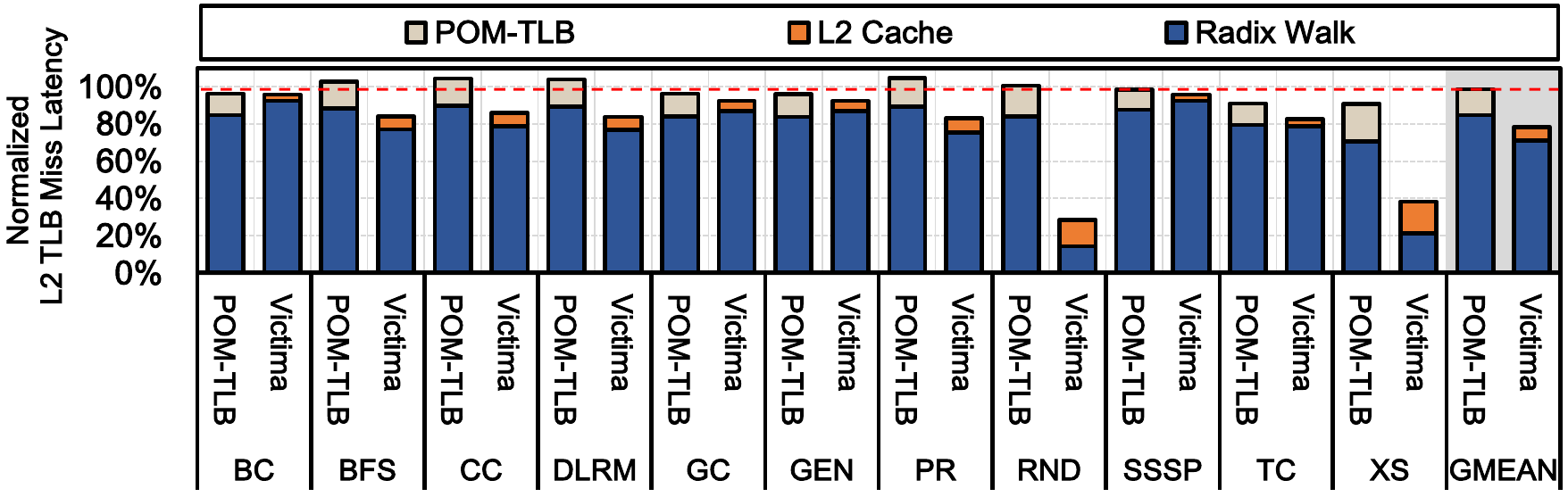}
    \vspace{-6mm}
    \caption{L2 TLB miss latency in POM-TLB and \system normalized to Radix.} 
    \label{fig:victima_translation_latency}
    \vspace{-4mm}
\end{figure}

Figure~\ref{fig:victima_translation_latency} shows the reduction in L2 TLB miss latency for POM-TLB and Victima over Radix. 
\system and POM-TLB respectively reduce L2 TLB miss latency by 22\% and 3\% over Radix. 
We observe that the latency of accessing POM-TLB nearly nullifies the potential performance gains of reducing PTWs. 
We conclude that \system delivers significant performance gains compared to all evaluated systems due to the reduction \konreve{in} the number of PTWs
which in turn leads to a reduction \konreve{in} the total L2 TLB miss latency.

\subsection{Diving Deeper into \system}


\subsubsection{Translation Reach.}
\begin{figure}[h]
    \vspace{-2mm}
    \centering
    \includegraphics[width=\linewidth]{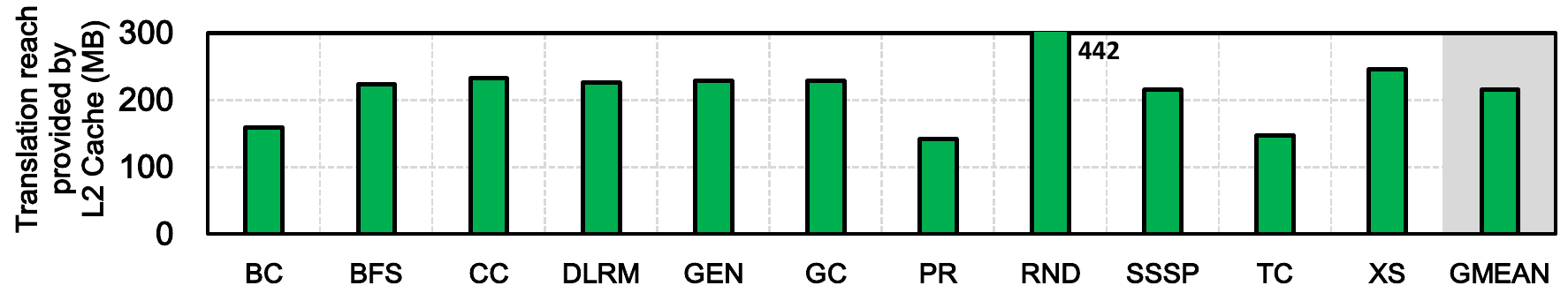}
    \vspace{-6mm}
    \caption{Translation reach provided by TLB blocks stored in L2 cache (assuming 4KB page size).}
    \label{fig:reach_victima}
    \vspace{-3mm}
\end{figure}

Figure~\ref{fig:reach_victima} shows the translation reach of a processor that uses \system \konreve{averaged across 500K execution epochs.}\footnote{\konreve{Each epoch consists of 1K instructions and we assume 4KB pages for simplicity.}} 
We observe that the average translation reach provided by \system is 36x larger (220 MBs) 
than the maximum reach offered by the L2 TLB of the baseline system that uses a two-level TLB hierarchy. 
This is due to the fact that each cache block can cover 32KBs (16MB) of memory while each L2 TLB block covers 4KB (2MB) 
per entry and the L2 cache typically has significantly more blocks than the L2 TLB (e.g., a 2MB cache has 21$\times$ \konreve{the blocks} of a 1.5K-entry L2 TLB).

\subsubsection{Reuse of TLB Blocks.}
\begin{figure}[h!]
    \vspace{-3mm}

    \centering
    \includegraphics[width=\linewidth]{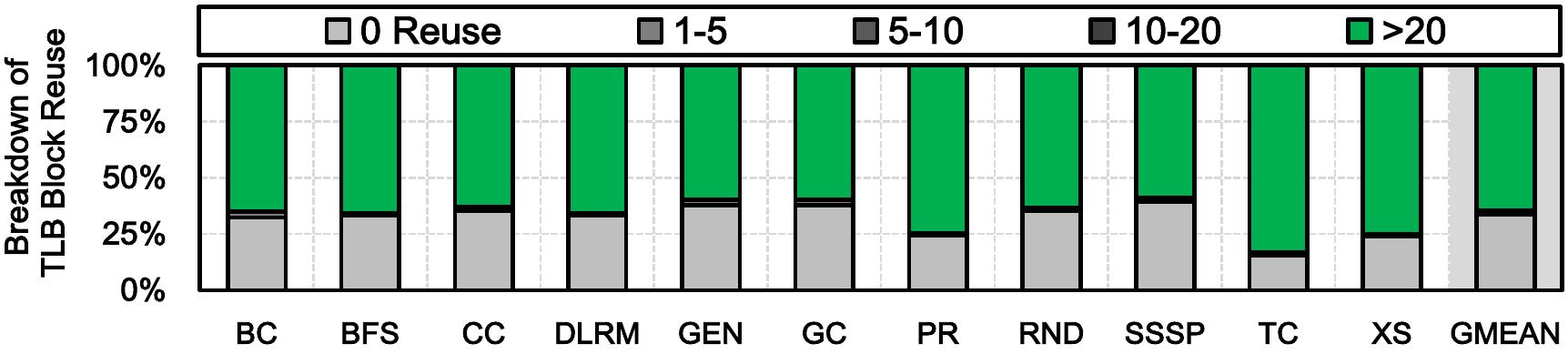}
    \vspace{-6mm}
    \caption{\konrevc{Reuse-level distribution of TLB blocks in L2 cache.}}
    \label{fig:reuse_tlb_blocks}
    \vspace{-4mm}

\end{figure}

Figure~\ref{fig:reuse_tlb_blocks} shows the reuse \konreve{distribution} of the TLB blocks in the L2 cache 
(we measure \konreve{a block's} reuse once the block gets evicted from the L2 cache). 
\konrevf{We observe that the majority of TLB blocks (65\%) experience high reuse (i.e., accessed more than 20 times 
before getting evicted from the L2 cache) due to (i) the accuracy of the PTW-CP (82\% average accuracy across all workloads) and 
(ii) the prioritization of the TLB blocks by the TLB-aware replacement policy used in the L2 cache. 
We conclude that Victima effectively utilizes underutilized L2 cache resources to store high-reuse TLB blocks.
In a system without Victima, accessing these TLB blocks would lead to high-latency PTWs.}

\subsubsection{Sensitivity to L2 Cache Size}

\begin{figure}[h]
    \vspace{-2mm}
    \centering
    \includegraphics[width=\linewidth]{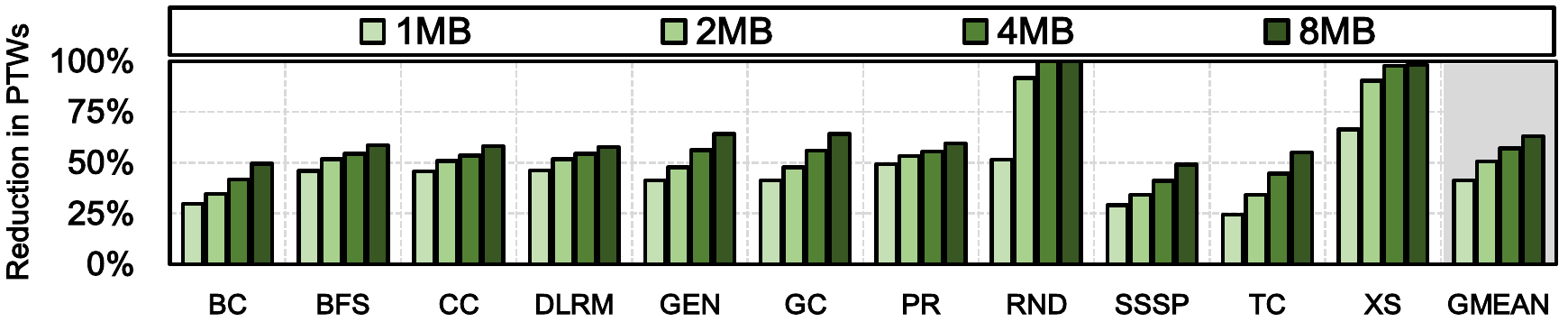}
    \vspace{-7mm}
    \caption{Victima's reduction in PTWs across different L2 cache sizes.}
    \label{fig:sensitivity_to_L2_cache_size}
    \vspace{-4mm}

\end{figure}

Figure~\ref{fig:sensitivity_to_L2_cache_size} shows the reduction in PTWs  achieved by Victima for \konrevd{four different L2 cache sizes}, ranging from 1MB up to 8MB. 
We \konrevd{observe} that Victima increasingly reduces PTWs with increasing L2 cache sizes. 
For the 8MB cache configuration, \system achieves the highest reduction \konrevd{in PTWs}, 63\% compared to Radix.
This can be attributed to the fact that a larger L2 data cache allows for caching more TLB blocks,
thereby increasing the translation reach of the processor.

\subsubsection{Sensitivity to L2 Cache Replacement Policy.}
\begin{figure}[h]
    \vspace{-2mm}

    \centering
    \includegraphics[width=\linewidth]{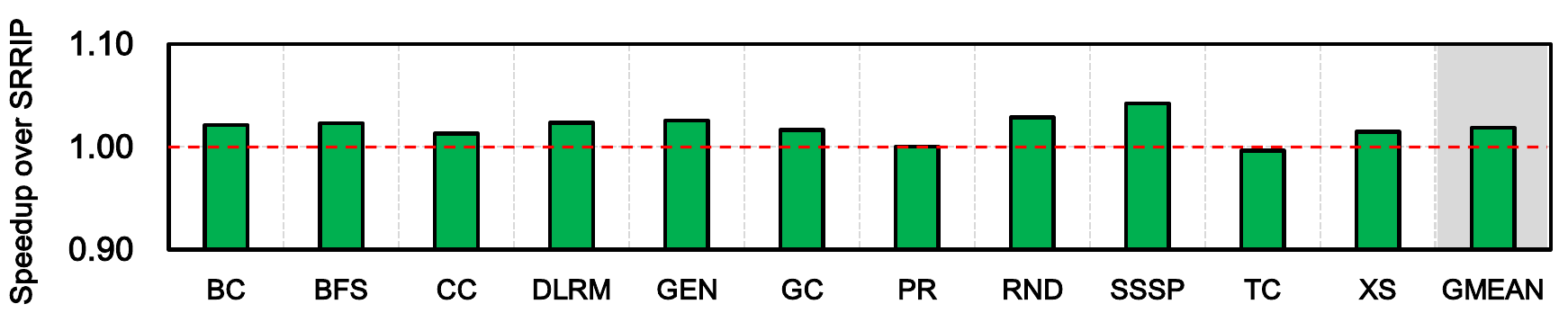}
    \vspace{-6mm}
    \caption{\konkanellorevb{\konreve{Performance improvement provided by Victima with \konrevs{the TLB-aware} SRRIP replacement policy over \konrevs{Victima} with TLB-agnostic SRRIP.}}}
    \label{fig:sensitivity_to_L2_cache_repl}
    \vspace{-3mm}
\end{figure}

\konkanellorevb{Figure~\ref{fig:sensitivity_to_L2_cache_repl} shows the performance of Victima when employing the TLB-aware SRRIP replacement policy at the L2 cache compared to employing a conventional TLB-agnostic SRRIP replacement policy.
We observe that employing the TLB-aware SRRIP leads to 1.8\% higher performance compared to the conventional SRRIP.}
We conlude that Victima can deliver high performance with both TLB-aware and TLB-agnostic replacement policies.

\subsection{Virtualized Environments}
\label{sec:virtualized_results}
\begin{figure}[h!]
    \vspace{-3mm}
    \centering
    \includegraphics[width=\linewidth]{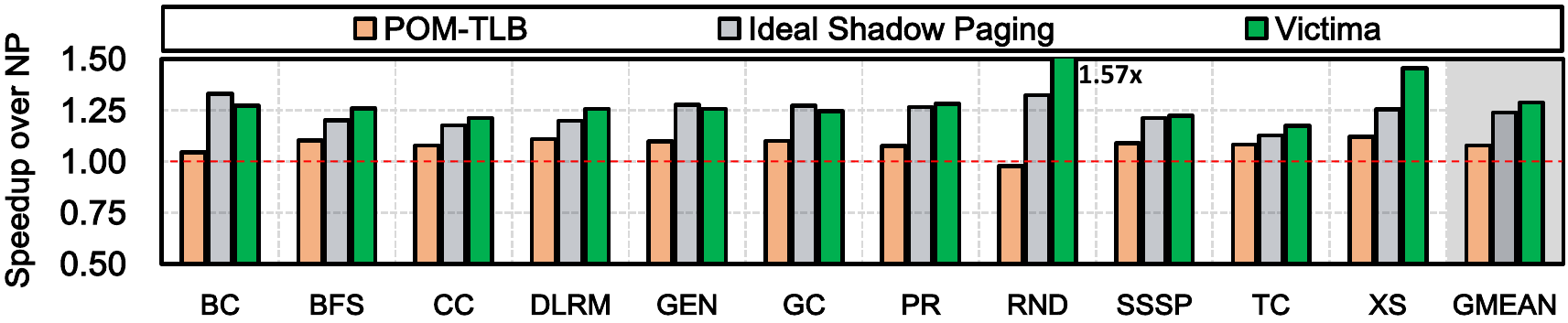}
    \vspace{-6mm}
    \caption{\konreve{Speedup provided by POM-TLB, I-SP and \system in a virtualized system with NP.}}
    \label{fig:speedup_virtualized_victima}
    \vspace{-1mm}

\end{figure}

Figure~\ref{fig:speedup_virtualized_victima} shows the execution time speedup of POM-TLB, I-SP and \system over \dbb{Nested Paging}, in a virtualized execution environment, across 11 workloads. 
We observe that \system outperforms \dbb{Nested Paging} on average by $\speedupoverbaselinevirt\%$, I-SP by $\speedupoverispvirt\%$, and POM-TLB by $\speedupoverpomtlbvirt\%$, across all workloads.  
To better understand the performance speedup achieved by \system, we examine the impact of \system on (i) the number of guest and host PTWs and (ii) \konreve{the L2 TLB miss latency and the nested TLB miss latency.}

\begin{figure}[h!]
    \centering
    \vspace{-1mm}
    \includegraphics[width=\linewidth]{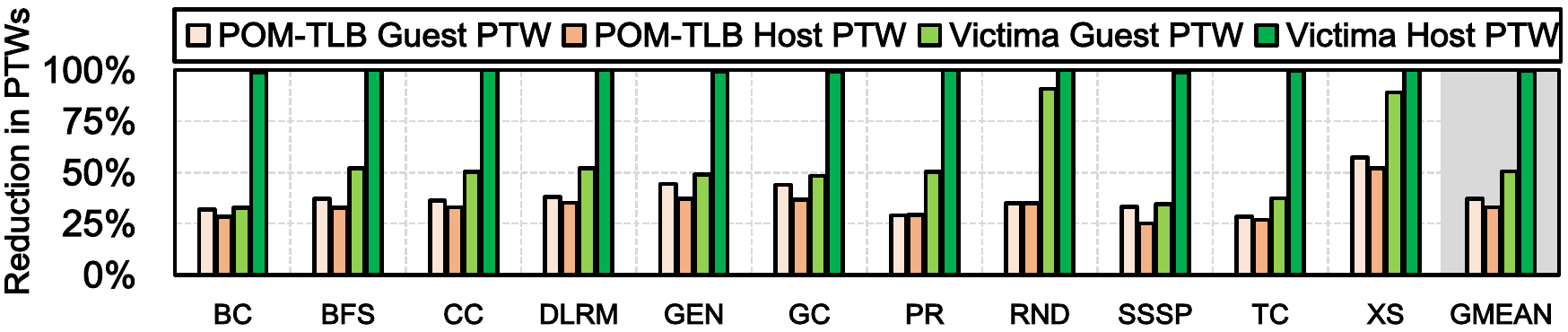}
    \vspace{-6mm}
    \caption{Reduction in host and guest PTWs \konreve{provided} by POM-TLB and Victima in a virtualized system with NP.}
    \vspace{-3mm}
    \label{fig:virtualized_ptw}
\end{figure}

Figure~\ref{fig:virtualized_ptw} shows the reduction in guest and host PTWs for all the configurations. 
We observe that \system leads to significant reductions in both guest PTWs ($50\%$) and host PTWs ($99\%$). 
The host PTW is the major bottleneck in NP, and \system almost eliminates it by caching nested TLB blocks inside the L2 cache. 

\begin{figure}[h!]
    \centering
    \includegraphics[width=\linewidth]{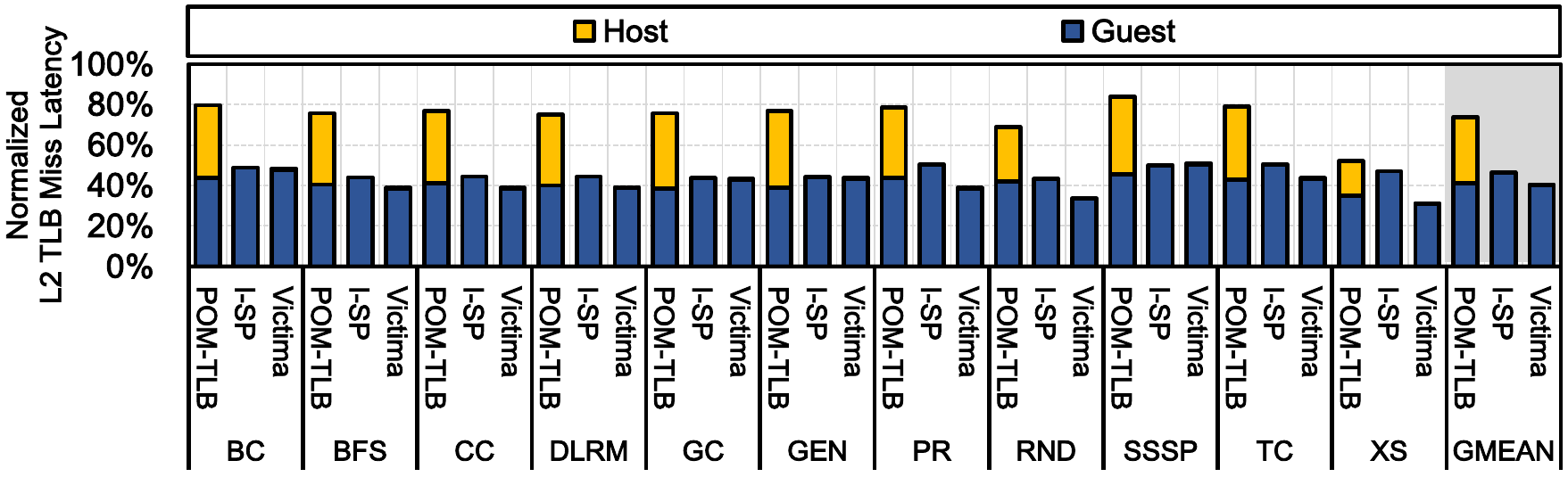}
    \vspace{-6mm}
    \caption{\konreve{L2 TLB miss latency in POM-TLB, I-SP and Victima normalized to NP.}}
    \label{fig:virtualized_norm_latency}\
    \vspace{-5mm}
\end{figure}

Figure~\ref{fig:virtualized_norm_latency} shows the L2 TLB miss latency for all the configurations normalized to NP. We observe that
Victima minimizes host PTW latency to as low as $1\%$ of the baseline while reducing the guest translation by 60\%, 6\% 
more than I-SP, which performs only four PT accesses to find out the guest-virtual to host-physical translation. 
We conclude that caching both nested and conventional TLB entries 
in the L2 cache allows Victima to achieve high performance in both native and virtualized environments.

\vspace{-2mm}
\section{Related Work}

To our knowledge, Victima is the first software-transparent mechanism that proposes 
caching TLB entries in the cache hierarchy to increase the translation reach of the processor. 
\konrevs{We have already comprehensively compared Victima to (i) systems that employ large hardware TLBs and large software-managed TLBs~\cite{pomtlbISCA2017} in native execution environments and (ii)
systems that employ nested paging~\cite{amdnested}, large software-managed TLBs~\cite{pomtlbISCA2017} and ideal shadow paging~\cite{vm25} in virtualized environments in \S\ref{sec:native_results} and \S\ref{sec:virtualized_results}.
In this section, we qualitatively compare Victima \konrevs{to other related} prior works that propose solutions to reduce address translation overheads.}

\head{Storing TLB Entries in On-chip Resources} Two prior GPU-focused works propose storing TLB entries in on-chip resources of GPUs [103, 104]. 
Jagadish et al.~\cite{gputlbreachMICRO2021} propose storing evicted L1 TLB entries in the underutilized instruction cache and the scratchpad memory of GPUs. 
Jaleel et al.~\cite{ducatiTACO2019} propose (i) storing TLB entries inside the LLC of a GPU as well as (ii) employing a large software-managed TLB that covers the whole system memory. 
Instead of inserting all evicted TLB entries into the on-chip resources as proposed in \cite{gputlbreachMICRO2021,ducatiTACO2019}, Victima employs a prediction mechanism (i.e., PTW-CP) that estimates the future cost 
of translating a virtual address and decides whether to insert or not the corresponding TLB entry inside the cache hierarchy. Thus, Victima (i) does not waste cache capacity for TLB entries 
that do \emph{not} benefit from being stored in the cache hierarchy and (ii) avoids using additional software structures to store translation metadata. Our evaluation results show that a scheme that 
combines (i) Victima and (ii) a large software-managed TLB that covers the whole main memory space, similar to the one used in DUCATI~\cite{ducatiTACO2019}, performs on average only 0.8\% better than Victima alone.

\head{Efficient TLBs and Page Walk Caches (PWCs)} Many prior works focus on reducing address translation overheads through efficient TLB and PWC designs\VMtlball.
Such techniques involve: (i) prefetching TLB and page table entries\VMtlbprefetching, 
(ii) TLB-specific replacement policies\VMtlbreplacementpolicy, (iii) employing software-managed TLBs\VMsoftwareTLB,
(iv) sharing TLBs across cores\VMtlblthree, (v) employing efficient PWCs\VMpwcs, and (vi) \konrevs{PT-aware cache management~\cite{pinningAccess2022,flataAsplos2022,consciousISPASS2022} (e.g., pinning PTEs in the LLC~\cite{pinningAccess2022}).}
Although such techniques may offer notable performance improvements, as the page table size increases, their effectiveness \konrevf{reduces}.
This is because they rely on
(i) the existing TLB hierarchy that is unable to accommodate the large number of TLB entries required by data-intensive applications or (ii) 
new hardware/software translation structures that pose a significant trade-off between performance and area/energy efficiency (\S\ref{sec:motivation}).
In contrast, Victima repurposes the \emph{existing} underutilized resources of the cache hierarchy to drastically increase 
\konrevf{address} translation reach and thus does \emph{not} require additional structures to store translation metadata.
For example, as we show in \S\ref{sec:motivation-stlb}, employing a software-managed TLB to back up the L2 TLB is not effective 
in native environments as the latency of the PTW is similar to the latency of accessing the software-managed TLB. 
In \S\ref{sec:virtualized_results} and \S\ref{sec:native_results}, we compare Victima against state-of-the-art software-managed TLB, POM-TLB~\cite{pomtlbISCA2017} and show that Victima
outperforms POM-TLB by \speedupoverpomtlb\% (\speedupoverpomtlbvirt\%) in native (virtualized) environments \konrevs{by} storing TLB entries in the 
high-capacity and low-latency L2 cache.

\head{Alternative Page Table Designs}
Various prior works focus on alternative page table designs\VMpagetable~to accelerate PTWs.
For example, Skarlatos et al.~\cite{elastic-cuckoo-asplos20} propose replacing the radix-tree-based page table with a Cuckoo hash table~\cite{fotakis} to parallelize accesses 
to the page table and reduce PTW latency. Park et al.~\cite{flataAsplos2022} propose a flat page table design in combination with a page-table-aware
replacement policy to \konrevs{reduce PTW latency}. Victima is \konrevf{complementary} to these techniques as it reduces PTWs while these techniques reduce PTW latency.

\head{Employing Large Pages} 
Many works propose hardware and software mechanisms for efficient and transparent support for pages of varying sizes\VMlargepages.  
For example, Ram et al.~\cite{tridentMICRO2021} propose harnessing memory resources to provide 1GB pages to applications in an \konrevs{application-transparent manner}. 
Guvenilir et al.~\cite{guvenilir2020tailored} propose modifications to the existing radix-based page table design to support a wide range of different page sizes. 
As we discuss in \S\ref{sec:l2-changes}, Victima is able to cache TLB entries for any page size and \konrevs{thus} is compatible with large pages.

\head{Contiguity-Aware Address Translation} 
\konrevs{Many prior works} \konrevs{enable and exploit} \konrevs{virtual-to-physical address} contiguity to perform \konrevs{low-latency} address translation\VMcontiguity. 
For example, in \cite{vm2}, \konrevs{the authors propose pre-allocating arbitrarily-large contiguous physical regions (10-100's of GBs) to drastically increase the translation reach for specific data structures of the application.}
Karakostas et al. \cite{karakostas2015} propose the use of multiple dynamically-allocated contiguous physical regions, called ranges, to provide efficient address translation for a small number of large memory objects used by the application.
Alverti et al.~\cite{chloe2020} propose an OS mechanism that enables efficient allocation of large contiguous physical regions.
These works can significantly increase translation reach, but, in general \konrevf{have two drawbacks: (i) they require system software modifications and (ii) their effectiveness heavily depends on the availability of free contiguous memory blocks.}
In contrast \konrevs{to these works}, Victima increases the translation reach of the processor
without requiring (i) contiguous physical memory allocations \konrevs{or} (ii) modifications \konrevs{to the system software}.

\head{Address Translation in Virtualized Environments}
\konrevs{Various works propose techniques to} reduce address translation overheads in virtualized environments\VMvirtualized.
For example, Ghandi et al. \cite{vm25} propose a hybrid address translation design for virtualized environments that combines shadow paging and nested paging.
In \S\ref{sec:virtualized_results}, we show that Victima is effective in virtualized environments and 
outperforms an ideal shadow paging design by \speedupoverispvirt\% \konrevs{by} storing both TLB \konrevs{entries} and nested TLB entries in the cache hierarchy.

\head{Virtual Caching \& Intermediate Address Spaces} Another class of works focuses on delaying address translation by using techniques such as 
virtual caching\VMvirtualcaching~and intermediate address spaces\VMintermediate. 
Virtually-indexed caches reduce address translation overheads by performing address translation only after a memory request 
misses in the LLC~\cite{basu2012, cekleov1997a, cekleov1997b, wood1986}. 
Hajinazar et al.~\cite{vbi} propose the use of virtual blocks mapped to an intermediate address space to delay
address translation until an LLC miss. 
\konrevf{Victima is orthogonal to these techniques and can operate \konrevs{with both (i)} virtually-indexed caches\footnote{Victima distinguishes between data blocks and TLB entries by using a tag bit in the cache block, regardless of whether the cache is virtually- or physically-indexed.} 
\konrevs{and (ii)} intermediate address spaces by storing TLB blocks with intermediate-to-physical address mappings in the cache hierarchy.}

\section{Conclusion}
Data-intensive workloads experience frequent and long-latency page table walks. 
\begin{sloppy}
This paper introduces \system, a software-transpa\-rent technique that stores TLB entries in the cache hierarchy to 
drastically increase the translation reach of the processor and thus reduces \konrevs{the occurence of page table walks}\end{sloppy}.
Our evaluation shows that \system provides significant performance improvements in both native and virtualized environments.
\system presents a practical opportunity to improve the performance of data-intensive workloads with small hardware changes, modest area and power overheads, 
and no modifications to software, by repurposing the underutilized resources of the \konrevs{cache hierarchy}.

\begin{acks}
  {
    We thank the anonymous reviewers of MICRO 2023 for their encouraging feedback.
    We thank the SAFARI Research Group members for providing a
    stimulating intellectual environment. We acknowledge the generous gifts from our industrial partners: Google, Huawei, Intel,
    Microsoft, and VMware. This work is supported in part by
    the Semiconductor Research Corporation and the ETH Future
    Computing Laboratory. 
  }
\end{acks}

 \bibliographystyle{unsrt}

\bibliography{refs}
\appendix

\section{Artifact Appendix} \label{sec:artifact}

\subsection{Abstract}

We implement Victima using the Sniper simulator~\cite{sniper}. 
In this artifact, we provide the source code of Victima and necessary instructions to reproduce its key performance results. 
We identify four key results to demonstrate \konrevf{and analyze Victima's novelty and effectiveness}: 
\begin{itemize}
    \item Execution time speedup and MPKI reduction achieved by increasing L2 TLB size and by using an L3 TLB.
    \item Accuracy, recall, precision and F1-score of the comparator-based PTW-Cost Predictor.
    \item Execution time speedup, PTW reduction and increase of address translation reach provided by Victima.
    \item Execution time speedup and PTW reduction in virtualized environments.
\end{itemize}

The artifact can be executed in any machine with a general-purpose CPU and $10$ GB disk space. 
However, we strongly recommend running the artifact on a compute cluster with \texttt{slurm}~\cite{slurm} support for bulk experimentation.

\subsection{Artifact Check-list (Meta-information)}

{\small
\begin{itemize}

  \item {\bf Compilation: } Container-based compilation.
  \item {\bf Data set: } Download traces using the supplied script.
  \item {\bf Run-time environment: } Docker-based environment.
  \item {\bf Metrics: } TLB MPKI, Speedup, Reuse and Translation Reach.
  \item {\bf Experiments: } Generate experiments using supplied scripts.
  \item {\bf How much disk space required (approximately)?: } $10$GB
  \item {\bf How much time is needed to prepare workflow (approximately)?: } $\sim 0.5$ hours. Mostly depends on the time to download traces.
  \item {\bf How much time is needed to complete experiments (approximately)?: } 8-10 hours using a compute cluster with $250$ cores.
  \item {\bf Publicly available?: } Yes.
  \item {\bf Archived (provide DOI)?: } \url{https://zenodo.org/record/8220613}
\end{itemize}
}
\subsection{Description}

\head{How to Access} The source code can be downloaded either from GitHub (\url{https://github.com/CMU-SAFARI/Victima}) or from Zenodo (\url{https://zenodo.org/record/8220613}).
\newline \\
\head{Hardware Dependencies} We use Docker/Podman images to execute the experiments. All container images support x86-64 architectures.
\begin{itemize}
    \item {The experiments were executed using \texttt{Slurm}~\cite{slurm}. \newline We strongly suggest executing the experiments using such an infrastructure for bulk experimentation. However, we provide support for non-slurm-based, native execution.}
    \item Each experiment takes $\approx$8-10 hours to finish and requires about $\approx$5-13GB of free memory (depends on the experiment).
    \item The workload traces require $\approx10$GB of storage space.
    \item {Hardware infrastructure used to run the experiments: \newline (i) Nodes:  Intel(R) Xeon(R) Gold 5118 CPU @ 2.30GH and \newline (ii) \texttt{Slurm}~\cite{slurm} version: slurm-wlm 21.08.5.}
\end{itemize}

\head{Software Dependencies} All container images are publicly available in Docker hub under the tags:
\begin{enumerate}
    \item Contains all the simulator dependencies \\
\shellcmd{docker.io/kanell21/artifact\_evaluation:victima }
    \item Contains all python dependencies to reproduce the results of Table 2 and create plots \\
\shellcmd{docker.io/kanell21/artifact\_evaluation:victima\_ptwcp\_v1.1 }
\end{enumerate}
\vspace{0.2cm}

To execute experiments with a container, we need the following software (all packages will be automatically downloaded using the provided scripts):
\vspace{0.2cm}
{\small
\begin{itemize}
    \item \texttt{Docker: \{docker-ce, docker-ce-cli,containerd.io\}}
    \item \texttt{Podman (instead of Docker)}
    \item \texttt{curl}
    \item \texttt{tar}
\end{itemize}
}

In our experiments we used the following packages:

{\small
\begin{itemize}
    \item \texttt{Docker version 20.10.23, build 7155243}
    \item \texttt{Podman 3.4.4}
    \item \texttt{curl 7.81.0}
    \item \texttt{tar (GNU tar) 1.34}
    \item \texttt{Kernel: 5.15.0-56-generic}
    \item \texttt{Dist: Ubuntu SMP 22.04.1 LTS (Jammy Jellyfish)}
\end{itemize}
}

\head{Data Sets} The Sniper traces required to evaluate Victima will be downloaded automatically using the supplied scripts. 
All traces are uploaded in Google Cloud Storage under this link: \url{https://storage.googleapis.com/traces_virtual_memory/traces_victima}


\subsection{Experiment Workflow}
This section describes steps to install all required software and execute necessary experiments. 
We recommend the reader to follow the README file to know more about each script used in this section.
\newline \\
\head{Installation} \label{sec:artifact_launching_experiments} The following \konrevf{command line} instructions will install all software packages for Docker or Podman.

\begin{enumerate}
    \item Specify Podman or Docker: \\
        \shellcmd{\textasciitilde/Victima\$ sh install\_container.sh podman | docker}
\end{enumerate}

\head{Launching Experiments}\label{sec:artifact_launching_experiments} The following script downloads all traces and launches all experiments 
required to reproduce the key results. The following command directly executes neural network inference to reproduce the results shown in Table 2. 
We \textbf{strongly} recommend using a compute cluster with \texttt{Slurm} support to efficiently launch experiments in bulk. We have set the maximum 
memory usage of each slurm job as 10GB and the maximum timeout as 3 days, in order to make sure that all experiments will run correctly.

\begin{enumerate}
    \item {To launch your experiments execute : \\
          \shellcmd{\textasciitilde/Victima\$ sh artifact.sh  -{}-slurm docker~\textbar~podman }}

\end{enumerate}

\head{Parse Results \& Generate Figures} All results are stored under ./results.
Execute the following command to:

\begin{enumerate}
\item  Parse the results of the experiments. 
\item  Generate Figures 5, 6, 8, 11, 20-21, 23-25, 27, 28. All figures can be found under:  \texttt{/path/to/Victima/plots/} 
\end{enumerate}
         
\shellcmd{\textasciitilde/Victima\$ sh ./scripts/produce\_plots.sh docker~\textbar~podman}

\section{Reusability using MLCommons}

We added support to evaluate Victima using the {MLCommons CM automation language}: \url{https://github.com/mlcommons/ck}.
Make sure you have installed CM. Follow the guide under \url{https://github.com/mlcommons/ck/blob/master/docs/installation.md} to install it.
Next, install reusable MLCommons automations and pull this repository via CM::

\vspace{0.2cm}
\shellcmd{cm pull repo mlcommons@ck \&\& cm pull repo CMU-SAFARI@Victima}

\vspace{0.2cm}

\noindent The CM scripts for Victima will be available under: \\
\texttt{/CM/repos/CMU-SAFARI@Victima/script/}. Perform the following steps to evaluate Victima with MLCommons:

\begin{enumerate}
    \item {\shellcmd{cm run script micro-2023-461:install\_dep \textbackslash  \\ --env.CONTAINER\_461="docker" }}

    \item {\shellcmd{cm run script micro-2023-461:run-experiments \textbackslash  \\--env.EXEC\_MODE\_461="--slurm" | native \textbackslash  \\
    --env.CONTAINER\_461="docker" | "podman"}}

    \item {\shellcmd{cm run script micro-2023-461:produce-plots \textbackslash  \\ --env.CONTAINER\_461="docker" | "podman"}}

\end{enumerate}
\subsection{Evaluation \& Expected Results}
The experiments evaluate (i) a system with different L2 TLB sizes, a system that employs an L3 TLB, POM-TLB~\cite{pomtlbISCA2017} and Victima in native execution environments as well as (ii) nested paging~\cite{amdnested}, POM-TLB~\cite{pomtlbISCA2017}, ideal shadow paging~\cite{vm25} and Victima in virtualized environments. 

\begin{itemize}

    \item Increasing the L2 TLB size up to 64K entries should lead to 4.0\% higher performance compared to the baseline system (Fig.~\ref{fig:l2-tlb-perf-opt}) and L2 TLB MPKI reduction from 39.4 to 24.3 (Fig.~\ref{fig:l2-tlb-mpki}). 
    Using an L3 TLB should lead to 2.9\% higher performance compared to the baseline system (Fig.~\ref{fig:l3tlb}).
    \item 92\% of L2 data blocks should experience a reuse of 0 and 8\% should experience a reuse higher than 1 (Fig.~\ref{fig:l2-reuse}).
    \item The comparator-based PTW-CP should achieve 89\% Recall, 82\% Accuracy, 73\% Precision and 80\% F1-Score (Table~\ref{tab:ptw_perf}). 
    \item Victima should outperform Baseline, POM-TLB, Optimistic L3 TLB 64K, Optimistic L2 TLB 64K, Optimistic L2 TLB 128K, by \speedupoverbaseline\%, 6.2\%, 4.4\%, 3.3\%, and 0.3\% respectively in native execution environments (Fig.~\ref{fig:speedup_victima}).
    \item Victima should reduce the number of PTWs by 50\% compared to the baseline system (Fig.~\ref{fig:victima_ptws}).
    \item Victima should provide 220 MBs of translation reach (Fig.~\ref{fig:reach_victima}).
    \item Victima should outperform Nested Paging, POM-TLB, and Ideal Shadow Paging by \speedupoverbaselinevirt\%, \speedupoverpomtlbvirt\%, and \speedupoverispvirt\% respectively in virtualized environments (Fig.~\ref{fig:speedup_virtualized_victima}).
    
\end{itemize}


\vspace{0.5mm}
\subsection{Methodology}

Submission, reviewing and badging methodology:
{\small
\begin{itemize}
  \item \url{https://www.acm.org/publications/policies/artifact-review-badging}
  \item \url{http://cTuning.org/ae/submission-20201122.html}
  \item \url{http://cTuning.org/ae/reviewing-20201122.html}
\end{itemize}
}

\end{document}